\documentclass[12pt]{article}
\pdfoutput=1

\usepackage[DIV13]{typearea}
\usepackage{indentfirst}
\RequirePackage{amsmath}
\RequirePackage{epsfig}
\RequirePackage{graphicx}
\usepackage{subfig}
\usepackage{multicol}
\usepackage{cite}
\usepackage[usenames,dvipsnames]{color}
\RequirePackage[colorlinks=true
,urlcolor=blue
,anchorcolor=blue
,citecolor=blue
,filecolor=blue
,linkcolor=blue
,menucolor=blue
,pagecolor=blue
,linktocpage=true
,pdfproducer=medialab
]{hyperref}

\newcommand{\email}[1]{\href{mailto:#1}{\tt #1}}

\numberwithin{equation}{section}

\newcommand{\la}{\lambda}
\newcommand{\eps}{\epsilon}
\newcommand{\vep}{\varepsilon}

\def\cY{{\cal Y}}
\def\cV{{\cal V}}
\def\cM{{\cal M}}
\def\cW{{\cal W}}
\def\cG{{\cal G}}
\def\cF{{\cal F}}
\def\cA{{\cal A}}

\newcommand{\be}{\begin{equation}}
\newcommand{\ee}{\end{equation}}
\newcommand{\beq}{\begin{equation}}
\newcommand{\eeq}{\end{equation}}
\newcommand{\bac}{\beq\begin{array}}
\newcommand{\eac}{\end{array}\eeq}
\newcommand{\ba}{\begin{array}}
\newcommand{\ea}{\end{array}}
\newcommand{\bea}{\begin{eqnarray}}
\newcommand{\eea}{\end{eqnarray}}

\newcommand{\tev}{\, {\rm TeV}}
\newcommand{\gev}{\, {\rm GeV}}
\newcommand{\mev}{\, {\rm MeV}}
\newcommand{\ps}{\, {\rm ps}}

\DeclareMathOperator{\RE}{Re}
\DeclareMathOperator{\IM}{Im}
\newcommand{\ord}{{\cal O}}
\newcommand{\Heff}{{\cal H}_\text{ eff}}

\newcommand{\hc}{\text{h.c.}}

\newcommand{\mean}[1]{\langle#1\rangle}

\newcommand{\ov}[1]{\overline{#1}}
\newcommand{\vcb}{|V_{cb}|}

\newcommand{\vub}{|V_{ub}|}

\newcommand{\vus}{|V_{us}|}

\begin{document}
\begin{titlepage}
\vspace*{-1cm}
\phantom{hep-ph/***}

\flushright
\hfil{TUM-HEP-821/11}
\hfill{FLAVOUR(267104)-ERC-6}\\

\vskip 1.5cm
\begin{center}
\mathversion{bold}
{\LARGE\bf Phenomenology of}\\[3mm]
{\LARGE\bf a Gauged $SU(3)^3$ Flavour Model}
\mathversion{normal}
\vskip .3cm
\end{center}
\vskip 0.5  cm
\begin{center}
{\large Andrzej J. Buras}~$^{a,b)}$,
{\large Maria Valentina Carlucci}~$^{a)}$,\\[2mm]
{\large Luca Merlo}~$^{a,b)}$,
{\large and Emmanuel Stamou}~$^{a,b,c)}$
\\
\vskip .7cm
{\footnotesize
$^{a)}$~Physik-Department, Technische Universit\"at M\"unchen, 
\\
James-Franck-Strasse, D-85748 Garching, Germany
\\
\vskip .1cm
$^{b)}$~
TUM Institute for Advanced Study, Technische Universit\"at M\"unchen, \\
Lichtenbergstrasse 2a, D-85748 Garching, Germany
\\
\vskip .1cm
$^{c)}$~Excellence Cluster Universe, Technische Universit\"at M\"unchen,\\
Boltzmannstrasse 2, D-85748 Garching, Germany
\vskip .5cm
\begin{minipage}[l]{.9\textwidth}
\begin{center} 
\textit{E-mail:} 
\email{andrzej.buras@ph.tum.de}, \email{maria.carlucci@ph.tum.de},\\[1mm] 
\qquad\,\,\email{luca.merlo@ph.tum.de}, \email{emmanuel.stamou@ph.tum.de}
\end{center}
\end{minipage}
}
\end{center}
\vskip 1.5cm
\begin{abstract}
We present an extensive analysis of $\Delta F=2$ observables 
and of $B\to X_s\gamma$ in the framework of a specific 
Maximally Gauged Flavour (MGF) model of 
Grinstein {\it et al.} including all relevant contributions, in particular tree-level heavy gauge boson exchanges whose effects are  studied in detail in the present paper for the first time. 
The model allows in principle for significant deviations from the Standard Model 
predictions for $\varepsilon_K$, $\Delta M_{B_{d,s}}$, mixing induced $CP$-asymmetries $S_{\psi K_S}$
and $S_{\psi \phi}$  and $B\to X_s\gamma$ decay. 
The tension between $\vep_K$ and $S_{\psi K_S}$ present in the SM can be removed by enhancing $|\vep_K|$ without modifying $S_{\psi K_S}$. In this case, we find that in this model i) the results 
for $S_{\psi \phi}$  and $B\to X_s\gamma$ turn out to be 
SM-like, ii) the {\it exclusive determination} of $\vub$ is favoured and most 
importantly iii) the values of $\Delta M_{B_d}$ and $\Delta M_{B_s}$ being strongly
correlated in this model with $\varepsilon_K$ turn  out to be much larger than the data for the central
values of input parameters: $\Delta M_{B_d}\approx0.75/ps$ and $\Delta M_{B_s}\approx27/ps$.
Therefore, from the present perspective, the model suffers from a serious 
$\varepsilon_K$--$\Delta M_{B_{d,s}}$ tension. However, this tension can be softened considering theoretical and parametric uncertainties and in particular the decrease of the weak decay constants. On the other side, the model can be strongly constrained considering the theoretically cleaner ratios $\Delta M_{B_d}/\Delta M_{B_s}$ and $BR(B^+\to\tau^+\nu)/\Delta M_{B_d}$ and we find that it is unable to remove simultaneously all the SM tensions on the data. Finally, we compare the pattern of flavour violation in MGF with selected extensions of the SM.
\end{abstract}
\end{titlepage}
\setcounter{footnote}{0}

\pdfbookmark[1]{Table of Contents}{tableofcontents}
\tableofcontents
\
%
%
\section{Introduction}
\label{sec:Intro}

The Standard Model (SM) of Particle Physics is successful in describing particles and their
electroweak and strong interactions, still, several aspects are problematic. In this paper,
we concentrate on the Flavour Problem.

The introduction of additional symmetries beyond the SM gauge group acting on the three fermion
generations can produce realistic mass hierarchies and mixing textures. The Lagrangian is
invariant under the gauge group of the SM and under the additional flavour symmetry at an
energy scale equal or higher than the electroweak one. Fermion masses and mixings arise
once these symmetries are broken, spontaneously or explicitly. Such flavour models differ from
each other in the nature of the symmetries and the symmetry breaking mechanism.
On the other hand, they all share the same top-down approach: the main goal is the explanation of
fermion masses and mixings by the introduction of flavour symmetries; only as a second step their phenomenological consistency with FCNC processes (sometimes) is investigated (see Refs.
\cite{Feruglio:2009hu,Merlo:2011hw} and references therein).

A bottom-up approach consists in first identifying a low-energy effective scheme in which
the contributions to FCNC observables are under control and subsequently in constructing 
high-energy models from which the effective description can be derived. The so-called Minimal
Flavour Violation (MFV) \cite{Chivukula:1987py,Hall:1990ac,Ciuchini:1998xy,Buras:2000dm} follows this
second approach. The fact that so far no evident deviations from the SM predictions have been found in any
flavour process observed in the hadronic sector \cite{Isidori:2010kg}, from rare decays in the
kaon and pion sector to $B$ decays at super$B$--factories, can be a sign that any physics beyond
the SM does not introduce significant new sources of flavour and CP violation with respect to the SM.
In Refs.~\cite{D'Ambrosio:2002ex,Cirigliano:2005ck,Davidson:2006bd,Alonso:2011jd}, this criterion has
been rigorously defined in terms of flavour symmetries, considering an effective operator description
within the SM. More in detail, restricted to the quark sector, the flavour symmetry coincides
with the symmetry of the SM Lagrangian in the limit of vanishing Yukawa couplings. This symmetry can
be written as the product of non-Abelian $SU(3)$ terms,
\begin{equation}
G_f=SU(3)_{Q_L}\times SU(3)_{U_R}\times SU(3)_{D_R}\,,
\label{FSMFV}
\end{equation}
and three additional $U(1)$ factors, that can be arranged to correspond to the Baryon number, the
Hypercharge and a phase transformation only on the right-handed (RH) down-type quarks. Interestingly,
only the non-Abelian terms of $G_f$ control the flavour structures of the quark mass-matrices,
while the $U(1)$ factors can only  be responsible for overall suppressions \cite{Alonso:2011jd}.
The $SU(2)_L$-doublet $Q_L$ and the $SU(2)_L$-singlets $U_R$ and $D_R$ transform under $G_f$ as 
\begin{equation}
Q_L\sim({\bf3},{\bf1},{\bf1})\,,\qquad\qquad
U_R\sim({\bf1},{\bf3},{\bf1})\,,\qquad\qquad
D_R\sim({\bf1},{\bf1},{\bf3})\,.
\end{equation}
In order to write the usual SM Yukawa terms,
\begin{equation}
\mathcal{L}_Y=\ov{Q}_L \cY_dD_RH+\ov{Q}_L \cY_uU_R\tilde{H}+\hc\,,
\end{equation}
where $\tilde{H}=i\tau_2H^*$, manifestly invariant under $G_f$, the Yukawa couplings are promoted
to dimensionless fields -- called spurions -- with non-trivial transformation properties under $G_f$:
\begin{equation}
 \cY_u\sim({\bf3},{\bf \ov{3}},{\bf1})\;,\qquad\qquad 
 \cY_d\sim({\bf3},{\bf1},{\bf\ov{3}})\;.
\label{YuYd}
\end{equation}
Following the MFV ansatz, quark masses and mixings arise once the electroweak symmetry is spontaneously
broken by the Higgs VEV, $\mean{H}=v/\sqrt2$ with $v=246$ GeV, and the spurion fields obtain the
values,
\begin{equation}
\cY_d=\dfrac{\sqrt2}{v}\left(
        \begin{array}{ccc}
           m_d  & 0 & 0 \\
            0 & m_s & 0 \\
            0 & 0 &  m_b \\
        \end{array}
	\right)\quad\text{and}\quad
\cY_u=\dfrac{\sqrt2}{v}\cV^\dag\left(
        \begin{array}{ccc}
           m_u  & 0 & 0 \\
            0 & m_c & 0 \\
            0 & 0 &  m_t \\
        \end{array}\right)\,,
\label{SpurionVEVsMFV}
\end{equation}
where $\cV$ is the unitary CKM matrix.

Recently, several papers \cite{Grinstein:2010ve,Feldmann:2010yp,Guadagnoli:2011id} appeared where
a MFV-like ansatz is implemented in the context of maximal gauge flavour (MGF) symmetries: in the limit
of vanishing Yukawa interactions these gauge symmetries are the largest non-Abelian ones allowed by the Lagrangian
of the model. The particle spectrum is enriched by
new heavy gauge bosons, carrying neither colour nor electric charges, and exotic fermions,
to cancel anomalies. Furthermore, the new exotic fermions give rise to the SM fermion
masses through a See-Saw mechanism, in a way similar to how the light left-handed (LH) neutrinos
obtain masses by the heavy RH ones. Moreover, the MFV spurions are promoted to scalar fields -- called flavons --
invariant under the gauge group of the SM, but transforming as bi-fundamental representations of
the non-Abelian part of the flavour symmetry. Once the flavons develop suitable VEVs, the SM fermion
masses and mixings are correctly described. Still, Refs.~\cite{Grinstein:2010ve,Feldmann:2010yp,Guadagnoli:2011id}
do not provide a natural mechanism for the specific structure of the flavon VEVs. This mechanism is highly model dependent, as discussed in Refs.~\cite{Feldmann:2009dc,Alonso:2011yg}, in contrast to the fermion and gauge sectors. Such scalar fields may have a phenomenological impact, but it is above the scope of the present analysis to provide a realistic explanation for the flavon VEV alignment and we will therefore not include these scalar contributions.

Even if this approach has some similarities to the usual MFV description, the presence of 
flavour-violating neutral gauge bosons and exotic fermions introduces modifications of the SM couplings and
tends to lead to dangerous contributions to FCNC processes mediated by the new heavy particles.
Consequently, the MGF framework goes beyond the standard MFV and a full phenomenological analysis 
of this NP scenario is mandatory to judge whether it is consistent with all available data.

In this paper we focus on the specific MGF realisation presented in Ref.~\cite{Grinstein:2010ve},
even if our analysis can be easily applied to other models with gauge flavour symmetries. In particular,
we extend the study performed in Ref.~\cite{Grinstein:2010ve} and point out that the parameter space
of such a model can be further constrained performing a full analysis on meson oscillations. 
The number of parameters is much smaller than in
other popular extensions of the SM and therefore it is not obvious
that the present tensions on the flavour data can be removed or at least softened. Indeed, we observe that the model, while solving the $\vep_K-S_{\psi K_S}$ tension, cannot simultaneously remove other SM flavour anomalies, which in some cases become even more pronounced.

Relative to Ref.~\cite{Grinstein:2010ve} the new aspects of our analysis are:
\begin{itemize}
\item[-]
In addition to new box-diagram contributions to $\Delta F=2$ processes, considered already in
Ref.~\cite{Grinstein:2010ve}, we perform a detailed analysis including the tree-level exchanges of new heavy flavour
gauge bosons. These diagrams generate LR operators that are strongly enhanced, by the
renormalisation group (RG) QCD running, relatively to the standard LL operators and could a priori be
very important. 
\item[-]
The impact of the new neutral current-current operators, arising from integrating out 
the heavy flavour gauge bosons, to the $\bar B\to X_s\gamma$ has been studied in Ref.~\cite{Buras:2011zb}
and we apply those results to the model.
\item[-]
We point out that for a value of $|V_{ub}|$ close to its 
determination from exclusive decays, i.e. in the ballpark of $3.5\times 10^{-3}$, the model can solve the present tension between $\varepsilon_K$ and $S_{\psi K_S}$. For slightly
larger values of $|V_{ub}|$, the model can still accommodate the considered observables within the errors,
but for the inclusive determination of $|V_{ub}|$ it suffers from tensions similar to the SM.
\item[-]
We scan over all NP parameters and present a correlated analysis of $\vep_K$, the mass differences $\Delta M_{B_{d,s}}$ the
$B^+\to\tau^+\nu$ and $\bar B\to X_s\gamma$ decays, the ratio $\Delta M_{B_d}/\Delta M_{B_s}$,
the mixing-induced CP asymmetries $S_{\psi K_S}$ and $S_{\psi\phi}$, and the $b$ semileptonic
CP-asymmetry $A^b_{sl}$.
\item[-]
We find that large corrections to the CP observables in the meson oscillations, $\vep_K$,
$S_{\psi K_s}$ and $S_{\psi\phi}$, are allowed. However, requiring $\vep_K$ to stay inside its $3\sigma$
error range, only small deviations from the SM values of $S_{\psi K_s}$ and $S_{\psi\phi}$
are allowed.
\item[-]
We find that requiring $\varepsilon_K$--$S_{\psi K_S}$ 
tension to be removed in this model implies 
the values of $\Delta M_{B_d}$ and $\Delta M_{B_s}$ to be significantly larger 
than the data. While the inclusion 
of theoretical and parametric uncertainties and in particular the decrease of the weak decay constants could soften this problem,  it appears from the present perspective that the model suffers from a serious $\varepsilon_K-\Delta M_{B{s,d}}$ tension.
\item[-]
We also investigate the correlation among two theoretically cleaner observables, $\Delta M_{B_d}/\Delta M_{B_s}$ and $BR(B^+\to\tau^+\nu)/\Delta M_{B_d}$. In this way, we strongly constrain the parameter space of the model and conclude that the tension in $BR(B^+\to\tau^+\nu)$, present already within the SM, is even increased.
\item[-]
We compare the patterns of flavour violation in this model
with those found in the original MFV, the MFV with the addition of flavour blind phases and MFV
in the left-right asymmetric framework.
\item[-]
As a by-product of our work we present a rather complete list of 
Feynman rules relevant for processes in the quark sector.
\end{itemize}

The structure of the paper is shown in the table of contents.

%
%

\section{The Model}
\label{sec:MGF}

In this section we summarise the relevant features of the MGF construction
presented in Ref.~\cite{Grinstein:2010ve}, dealing only with
the quark sector. The flavour symmetry is that of eq.~(\ref{FSMFV}), but it is gauged.
The spectrum is enriched by the corresponding flavour gauge bosons
and by new exotic quarks, necessary to cancel the anomalies: in particular the
new quarks are two coloured RH $SU(3)_{Q_L}$-triplets, one
LH $SU(3)_{U_R}$-triplet and one LH $SU(3)_{D_R}$-triplet. 
In table \ref{tab.GRV}, we list all the fields present in the theory and
their transformation properties under the gauge groups.

\begin{table}[ht]
\begin{center}
\begin{tabular}{|c||cccc||cccc|cc|}
\hline
&&&&&&&&&&\\[-4mm]
 & $Q_L$ & $U_R$ & $D_R$ & $H$ & $\Psi_{u_R}$ & $\Psi_{d_R}$ & $\Psi_{u_L}$ & $\Psi_{d_L}$ & $Y_u$ & $Y_d$ \\[2mm]
\hline
\hline
&&&&&&&&&&\\[-4mm]
$SU(3)_c$ & $\bf3$ & $\bf3$ & $\bf3$ & $\bf1$ & $\bf3 $ & $\bf3 $ & $\bf3 $ & $\bf3 $ & $\bf1$ & $\bf1$\\[2mm]
$SU(2)_L$ & $\bf2$ & $\bf1$ & $\bf1$ & $\bf2$ & $\bf1$ & $\bf1$ & $\bf1$ & $\bf1$ & $\bf1$ & $\bf1$ \\[2mm]
$U(1)_Y$ & $+^1/_6$ & $+^2/_3$ & $-^1/_3$ & $+^1/_2$ & $+^2/_3$ & $-^1/_3$ & $+^2/_3$ & $-^1/_3$ & $0$ & $0$ \\[2mm]
\hline
&&&&&&&&&&\\[-4mm]
$SU(3)_{Q_L}$ & $\bf3$ & $\bf1$ & $\bf1$ & $\bf1 $ & $\bf3$ & $\bf3$ & $\bf1 $ & $\bf1$ & $\bf\ov3$ & $\bf\ov3$ \\[2mm]
$SU(3)_{U_R}$ & $\bf1$ & $\bf3$ & $\bf1$ & $\bf1$ & $\bf1$ & $\bf1$ & $\bf3$ & $\bf1$ & $\bf3$ & $\bf1$ \\[2mm]
$SU(3)_{D_R}$ & $\bf1$ & $\bf1$ & $\bf3$ & $\bf1$ & $\bf1$ & $\bf1$ & $\bf1$ & $\bf3$ & $\bf1$ & $\bf3$\\[2mm]
\hline
\end{tabular}
\end{center}
\caption{\it The transformation properties of the fields under the SM and flavour gauge symmetries.}
\label{tab.GRV}
\end{table}

With this matter content, the most general renormalisable Lagrangian invariant
under the SM and flavour gauge groups can be divided into three parts:
\begin{equation}
\mathcal{L}=\mathcal{L}_{kin} + \mathcal{L}_{int} - V\left[H, Y_u, Y_d \right] \,.
\label{Lagrangian}
\end{equation}
The first one, $\mathcal{L}_{kin}$, contains the kinetic terms of all the fields and
the couplings of fermions and scalar bosons to the gauge bosons. 
The covariant derivative entering $\mathcal{L}_{kin}$ accounts for SM 
gauge boson-fermion interactions and
additional flavour interactions involving new gauge bosons and fermions:
\begin{equation}\label{eq:covariantderivatives}
D_{\mu} \supset \sum_{f=Q,U,D} i\, g_f\, N_f\, (A_f)_{\mu}\,,
\qquad\qquad\qquad (A_f)_{\mu}\equiv\sum_{a=1}^8 (A_f^a)_\mu\dfrac{\la_{SU(3)}^a}{2}\,,
\end{equation}
where $g_f$ are the flavour gauge coupling constants, $N_f$ the quantum numbers,
$A_f^a$ the flavour gauge bosons and $\la_{SU(3)}^a$ the Gell-Mann matrices.

The second term in eq.~(\ref{Lagrangian}), $\mathcal{L}_{int}$, contains the quark mass
terms and the quark-scalar interactions:
\begin{equation}\label{eq:lagrangian}
\begin{split}
\mathcal{L}_{\text{int}} =&\;\; \lambda_u\, \ov{Q}_L \tilde H\, \Psi_{u_R}+\lambda_u'\ov{\Psi}_{u_L} Y_u\, \Psi_{u_R} + M_u\, \ov{\Psi}_{u_L} U_R+ \\[2mm]
& + \lambda_d\, \ov{Q}_L H\, \Psi_{d_R}+\lambda_u'\ov{\Psi}_{d_L} Y_d\, \Psi_{d_R} + M_d\, \ov{\Psi}_{d_L} D_R + \hc\,,
\end{split}
\end{equation}
where $M_{u,d}$ are universal mass parameters and $\lambda^{(\prime)}_{u,d}$ are
universal coupling constants that can be chosen real, through a redefinition of
the fields.

The last term in eq.~(\ref{Lagrangian}), $V \left[H, Y_u, Y_d \right]$, is the
scalar potential of the model, containing the SM Higgs and the flavons $Y_{u,d}$.
The mechanisms of both electroweak and flavour symmetry breaking arise
from the minimisation of this scalar potential. It has not been explicitly constructed
in Ref.~\cite{Grinstein:2010ve} and it is beyond the scope of the present paper to
provide such a scalar potential 
(see Ref.~\cite{Alonso:2011yg} for a recent
analysis). Therefore, we  assume that
the spontaneous breaking of the electroweak symmetry proceeds as in the SM through
the Higgs mechanism and that the spontaneous flavour symmetry breaking is driven by the flavon fields $Y_{u,d}$ which develop the following VEVs:
\begin{equation}
\mean{Y_d}=\hat{Y}_d\,,\qquad\qquad\mean{Y_u}=\hat{Y}_u\,V\,. 
\label{eq:flavonvev}
\end{equation}
Here $\hat{Y}_{u,d}$ are diagonal $3\times3$ matrices and $V$ is a
unitary matrix. We emphasise that, despite the similarity to
eq.~(\ref{SpurionVEVsMFV}) of MFV, the matrix $V$ is not the CKM matrix
and the vacuum expectation values $\left\langle Y_{u,d} \right\rangle$ do not
coincide with the SM Yukawa matrices. This is illustrated by moving to
the fermion-mass eigenbasis. In what follows we focus on the up-quark sector, but
analogous formulae can also be written for the down-quark sector. The LH and RH
up-quarks mix separately giving rise to SM up-quarks $u^i_{R,L}$ and
exotic up-quarks $u^{\prime i}_{R,L}$:
\begin{equation}
\begin{pmatrix} 
		u^i_{R,L} \\[2mm] 
		u^{\prime i}_{R,L} 
\end{pmatrix} = 
\begin{pmatrix}
c_{u_{(R,L)i}} & -s_{u_{(R,L)i}} \\[2mm] 
s_{u_{(R,L)i}} & c_{u_{(R,L)i}} 
\end{pmatrix} 
\begin{pmatrix} 
U^i_{R,L} \\[2mm]
\Psi^i_{u_{R,L}}
\end{pmatrix}\,,
\end{equation}
where $c_{u_{(R,L)i}}$ and $s_{u_{(R,L)i}}$ are cosines and sines, respectively.
Denoting with $m_{f^i}$ the mass of the up-type $f^i=\{u^i,\,u^{\prime i}\}$ quark, what follows is a
direct inverse proportionality between $m_{u^i}$ and $m_{u^{\prime i}}$:
\begin{equation}
m_{u^i}\,m_{u^{\prime i}}=M_u\,\lambda_u\,\dfrac{v}{\sqrt2}\,.
\label{SeeSawMasses}
\end{equation}
We can express these masses in terms of the flavour symmetry breaking parameters:
\begin{equation}
m_{u^i}=\dfrac{s_{u_{Ri}}\,s_{u_{Li}}}{c^2_{u_{Ri}}-s^2_{u_{Li}}}\lambda'_u(\hat{Y}_{u})_i\,,\qquad\qquad
m_{u^{\prime i}}=\dfrac{c_{u_{Ri}}\,c_{u_{Li}}}{c^2_{u_{Ri}}-s^2_{u_{Li}}}\lambda'_u(\hat{Y}_{u})_i\,,
\end{equation}
where a straightforward calculation gives
\begin{equation}
s_{u_{Li}} =\sqrt{\dfrac{m_{u^i}}{M_u}\left\vert\dfrac{\lambda_u\,v\,m_{u^{\prime i}}-\sqrt2\,M_u\, m_{u^i}}{\sqrt2\,\left(m^{2}_{u^{\prime i}}-m^2_{u^i}\right)}\right\vert}
\,,\quad\quad
s_{u_{Ri}} =\sqrt{\dfrac{m_{u^i}}{\lambda_u\,v}\left\vert\dfrac{\sqrt2\,M_u\,m_{u^{\prime i}}-\lambda_u\,v\, m_{u^i}}{m^{2}_{u^{\prime i}}-m^2_{u^i}}\right\vert}\,.
\label{FormulaSin1}
\end{equation}

These results are exact and valid for all quark generations. However, taking
the limit $m_{u^{\prime i}}\gg m_{u^i}$, we find simple formulae that
transparently expose the behaviour of the previous expressions. In this limit we
find
\begin{align}
&m_{u^i} \approx  \dfrac{v}{\sqrt{2}} \dfrac{\lambda_u\, M_u}{\lambda'_u\, (\hat{Y}_u)_i}\,,\qquad\qquad
&&m_{u^{\prime i}}\approx \lambda'_u\, (\hat{Y}_u)_i\,,\\
&s_{u_{Li}}\approx \sqrt{\dfrac{m_{u^i}}{m_{u^{\prime i}}}\dfrac{\lambda_u\,v}{\sqrt2\,M_u}}\,,\qquad\qquad
&&s_{u_{Ri}}\approx \sqrt{\dfrac{m_{u^i}}{m_{u^{\prime i}}}\dfrac{\sqrt2\,M_u}{\lambda_u\,v}}\,,
\label{FormulaSin2}
\end{align}
as it is in the usual see-saw scheme in the limit of $(\hat{Y}_u)_i\gg M_u\,,v$.
These simplified relations are valid for all the fermions, apart from the top-quark
for which the condition $m_{t'}\gg m_t$ is not satisfied and large corrections
to eq.~(\ref{FormulaSin2}) are expected.

From eq.~(\ref{FormulaSin2}) we see that to reproduce the
correct SM quark spectrum, $\hat{Y}_u$ must have an inverted hierarchy
with respect to the SM Yukawas.\\
 
The presence of new exotic quarks has a relevant impact on the SM couplings. Indeed, the charged current-current interactions including SM and heavy quarks are governed by a $6\times6$ matrix which is
constructed from the unitary $3\times 3$ matrix $V$ of eq.~(\ref{eq:flavonvev})
and the $c_{u_{Li}}$, $c_{d_{Li}}$, $s_{u_{Li}}$ and $s_{d_{Li}}$ with ($i=1,2,3$)
introduced above. Adopting a matrix notation, the non-unitary $3\times3$ matrices
\begin{equation}\label{CKM1}
c_{u_L}\,V\,c_{d_L}\,,\quad\quad 
s_{u_L}\,V\,s_{d_L}
\end{equation}
describe the charged ($W^+$) current-current interactions within the
light and heavy systems, respectively. The analogous matrices
\begin{equation}\label{CKM2}
c_{u_L}\,V\,s_{d_L}\,,\quad\quad 
s_{u_L}\,V\,c_{d_L}
\end{equation}
describe the charged current-current interactions between
light and heavy fermions. In this notation, $c_{{u,d}_L}$ and $s_{{u,d}_L}$ are
diagonal matrices, whose entries are $c_{{u,d}_{Li}}$ and $s_{{u,d}_{Li}}$, respectively.
Moreover, we point out that in the no-mixing limit, $c_{{u,d}_{Li}}\to1$ and
$s_{{u,d}_{Li}}\to0$, the (non-unitary) $6\times6$ matrix reduces to
\begin{equation}
\begin{pmatrix} 
  V & 0 \\[2mm]
  0 & 0 
\end{pmatrix}\,.
\end{equation}
In this case the CKM matrix coincides with the unitary matrix $V$. As soon as the mixing
is switched on, the CKM is modified to include $c_{{u,d}_L}$, which
breaks unitarity. However, these deviations from unitarity are quite small (see sec.~\ref{sec:ckmmatrix}).
Moreover no new CP violating phases appear in the resulting CKM matrix. At first sight their absence implies 
no impact of new contributions to the CP-violating observables
$S_{\psi K_s}$ and $S_{\psi\phi}$. However, this is not the case due to the modification of the CKM
matrix and the presence of flavour gauge bosons. In this respect, this framework does differ from
the original MFV of Ref.~\cite{D'Ambrosio:2002ex}.

A consequence of the modification of the CKM matrix is the breaking of the GIM
mechanism if only SM quarks are considered in loop-induced processes. However,
once also the exotic quarks are included the GIM mechanism is recovered. We return to this
issue in sec.~\ref{sec:deltaF2properties}.

The interactions with the $Z$ boson and the Higgs field are
modified too. Their effects have been already discussed in
Ref.~\cite{Grinstein:2010ve} and it turned out that the largest constraint comes
from the modified $Z\,b\,\bar b$ coupling.\\

Once the flavour symmetry is spontaneously broken by the flavon VEVs, the flavour
gauge bosons acquire masses and mix among themselves. Using the vector notation for
the flavour gauge bosons,
\begin{equation}
\chi = \left( A_Q^1, \ldots, A_Q^8, A_U^1, \ldots, A_U^8, A_D^1, \ldots, A_D^8 \right)^T\,,
\end{equation}
the corresponding mass Lagrangian reads
\begin{equation}
\mathcal{L}_{\text{mass}} = \dfrac{1}{2}\,
\chi^T\,\cM_A^2\,\chi\,\qquad\text{with}\qquad
\cM_A^2 = \begin{pmatrix} 
M^2_{QQ} & M^2_{QU} & M^2_{QD} \\[1mm]
M^2_{UQ} & M^2_{UU} & 0 \\[1mm] 
M^2_{DQ} & 0 & M^2_{DD} 
\end{pmatrix}\,,
\label{eq:massmatrix}
\end{equation}
and
\begin{equation}
\begin{aligned}
\left( M^2_{QQ} \right)_{ab} = & \frac{1}{4}\, g_Q^2\, \text{Tr} \left[ \left\langle Y_u \right\rangle \left\{ \lambda_{SU(3)}^a,\lambda_{SU(3)}^b \right\} \left\langle Y_u \right\rangle^{\dag} + \left\langle Y_d \right\rangle \left\{ \lambda_{SU(3)}^a,\lambda_{SU(3)}^b \right\} \left\langle Y_d \right\rangle^{\dag} \right] \\
\left( M^2_{UU} \right)_{ab} = & \frac{1}{4}\, g_U^2\, \text{Tr} \left[ \left\langle Y_u \right\rangle \left\{ \lambda_{SU(3)}^a,\lambda_{SU(3)}^b \right\} \left\langle Y_u \right\rangle^{\dag} \right] \\
\left( M^2_{DD} \right)_{ab} = & \frac{1}{4}\, g_D^2\, \text{Tr} \left[ \left\langle Y_d \right\rangle \left\{ \lambda_{SU(3)}^a,\lambda_{SU(3)}^b \right\} \left\langle Y_d \right\rangle^{\dag} \right] \\
\left( M^2_{QU} \right)_{ab} = & \left( M^2_{UQ} \right)_{ba} = - \frac{1}{2}\, g_Q\, g_U\, \text{Tr} \left[ \lambda_{SU(3)}^a \left\langle Y_u \right\rangle^{\dag} \lambda_{SU(3)}^b \left\langle Y_u \right\rangle \right] \\
\left( M^2_{QD} \right)_{ab} = & \left( M^2_{DQ} \right)_{ba} = - \frac{1}{2}\, g_Q\, g_D\, \text{Tr} \left[ \lambda_{SU(3)}^a \left\langle Y_d \right\rangle^{\dag} \lambda_{SU(3)}^b \left\langle Y_d \right\rangle \right]\,.
\end{aligned}
\label{FGBmasses}
\end{equation}
In general, the diagonalisation of this mass-matrix is only numerically possible;
for the rest of the paper we shall indicate
with $\hat{\cM}^2_A$ the diagonal
matrix of the gauge boson mass eigenstates $\hat{A}^m$, where $m=1,\ldots,24$, and
with $\cW(\hat A^m,\,A_f^a)$, where $f=\{Q,\,U,\,D\}$ and $a=1,\ldots,8$, the
transformation to move from the flavour-basis to the mass-basis (see App.~\ref{app:coup2flavour}).

%
%

\section{\texorpdfstring{\mathversion{bold}$\Delta F=2$\mathversion{normal} Transitions}{Delta F=2 Transitions}}
\label{sec:DeltaF2}

\subsection{Effective Hamiltonian}

In the model in question the effective Hamiltonian for 
$\Delta F=2$ observables with external down-type quarks consists at the leading order in  weak and flavour-gauge 
interactions of two parts:
\begin{itemize}
\item[-]
Box-diagrams with SM $W$-boson and up-type quark exchanges. Due to the mixing among
light and heavy quarks, there are three different types of such diagrams:
with light quarks only, with heavy quarks only or with
both light and heavy quarks running in the box, as shown in
Fig.~\ref{fig:deltaF2box}. 
\begin{figure}[h!]
  \begin{center}
    \includegraphics{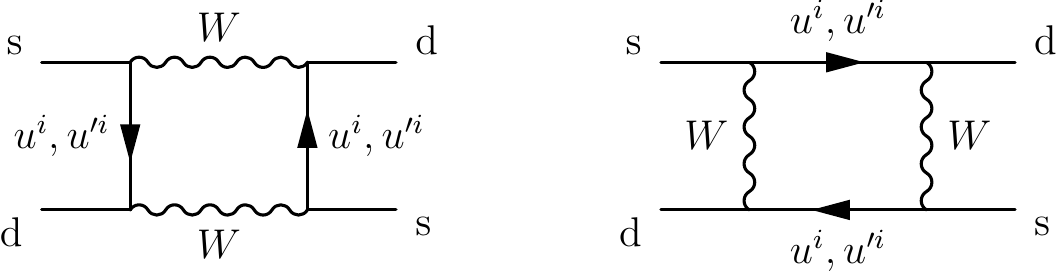}
  \end{center}
  \caption{\it The box-diagrams contributing to $K^0-\bar{K}^0$ mixing. Similarly for
  $B^0_q-\bar{B}^0_q$ mixing.}
  \label{fig:deltaF2box}
\end{figure}
If only exchanges of SM quarks are considered, the GIM mechanism is broken in these contributions.
It is recovered when also the exchanges of heavy quarks are taken into account.
\item[-]
The tree-level contributions from heavy gauge boson exchanges of
Fig.~\ref{fig:deltaF2tree}, that generate new neutral current-current operators, which violate flavour.
\begin{figure}[h!]
  \begin{center}
    \includegraphics{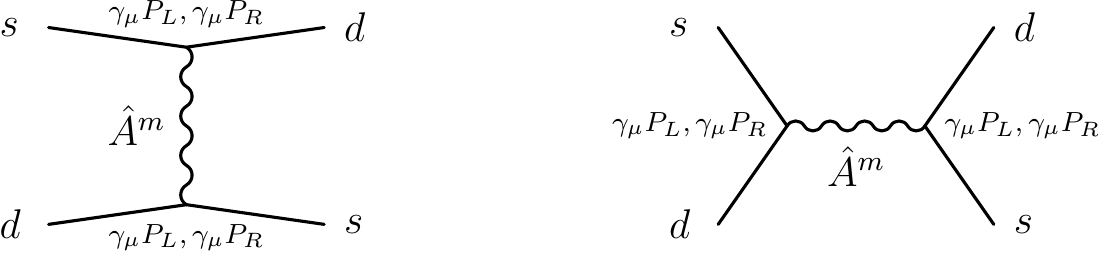}
  \end{center}
  \caption{\it The tree-diagrams contributing to $K^0-\bar{K}^0$ mixing. Similarly, for
  $B^0_q-\bar{B}^0_q$ mixing. $\hat{A}^m$ is a flavour gauge boson mass eigenstate.}
  \label{fig:deltaF2tree}
\end{figure}
\end{itemize} 

In principle one could consider box-diagrams with flavour-violating neutral 
heavy boson exchanges but they are negligible with respect to the tree-level contributions.

The effective Hamiltonian for $\Delta F=2$ transitions can then be written 
in a general form as 
\begin{equation}\label{Heff-general}
\Heff^{\Delta F=2} =\frac{G_F^2\,M^2_{W}}{4\pi^2}
\sum_{u^i} C_i(\mu)Q_i,
\end{equation}
where $M_W$ is the mass of the $W$-boson, $Q_i$ are the relevant operators for
the transitions, that we list below, and $C_i(\mu)$ their Wilson coefficients
evaluated at a scale $\mu$, which will be specified in the next section.

While in the SM only one operator contributes to each $\Delta F=2$ transition, i.e. $Q_1^{\rm VLL}(M)$ in the list of eq.~(\ref{normalM}), in
the model in question there are more dimension-six operators. In the absence of
flavon exchanges, the relevant operators for the $M^0$--$\bar{M}^0$ ($M=K,B_d,B_s$) systems 
are \cite{Buras:2000if}:
\begin{equation}
\begin{aligned}
Q_1^{\rm VLL}(K) &= (\bar{s}^{\alpha} \gamma_{\mu}    P_L d^{\alpha})
              (\bar{s}^{ \beta} \gamma^{\mu}    P_L d^{ \beta})\,,\qquad\quad
&Q_1^{\rm VLL}(B_q) &= (\bar{b}^{\alpha} \gamma_{\mu}    P_L q^{\alpha})
              (\bar{b}^{ \beta} \gamma^{\mu}    P_L q^{ \beta})\,,\\[2mm]               
Q_1^{\rm VRR}(K) &= (\bar{s}^{\alpha} \gamma_{\mu}    P_R d^{\alpha})
              (\bar{s}^{ \beta} \gamma^{\mu}    P_R d^{ \beta})\,,\qquad\quad
&Q_1^{\rm VRR}(B_q) &= (\bar{b}^{\alpha} \gamma_{\mu}    P_R q^{\alpha})
              (\bar{b}^{ \beta} \gamma^{\mu}    P_R q^{ \beta})\,,\\[2mm]  
Q_1^{\rm LR}(K) &=  (\bar{s}^{\alpha} \gamma_{\mu}    P_L d^{\alpha})
              (\bar{s}^{ \beta} \gamma^{\mu}    P_R d^{ \beta})\,,\qquad\quad
&Q_1^{\rm LR}(B_q) &=  (\bar{b}^{\alpha} \gamma_{\mu}    P_L q^{\alpha})
              (\bar{b}^{ \beta} \gamma^{\mu}    P_R q^{ \beta})\,,\\[2mm]   
Q_2^{\rm LR}(K) &=  (\bar{s}^{\alpha}                 P_L d^{\alpha})
              (\bar{s}^{ \beta}                 P_R d^{ \beta})\,,\qquad\quad
&Q_2^{\rm LR}(B_q) &=  (\bar{b}^{\alpha}                 P_L q^{\alpha})
              (\bar{b}^{ \beta}                 P_R q^{ \beta})\,.              
\end{aligned}
\label{normalM}
\end{equation}
where $P_{L,R} = (1\mp \gamma_5)/2$. 

In the next section, we collect the Wilson coefficients of these operators separating the
contributions from box-diagrams and from the tree-level heavy gauge boson exchanges
so that 
\begin{equation}
C^{(M)}_i=\Delta^{(M)}_{\rm Box}\,C_i+\Delta^{(M)}_{\rm A}C_i\,,
\end{equation}
where $M=K,\,B_d,\,B_s$.

\subsection{Wilson Coefficients from Box-Diagrams}
Keeping in mind the discussion around eqs.~(\ref{CKM1}) and (\ref{CKM2}) 
we introduce the mixing parameters:
\begin{equation}
\lambda_i(K)=V_{is}^{*}V_{id},\qquad 
\lambda_i(B_q)=V_{ib}^{*}V_{iq},
\end{equation}
where $q=d,s$ and $V$ is not the CKM matrix but the unitary matrix of eq.~\eqref{eq:flavonvev}.

Calculating the usual box-diagrams but including also contributions 
from heavy fermions (see Fig.~\ref{fig:deltaF2box}) 
and corrections to $W$-quark vertices according to the Feynman rules
in App.~\ref{app:coup2smgauge} we find the following contributions
to the Wilson coefficients relevant for the $K^0-\bar K^0$ system at the matching
scale $\mu_t$  in the ballpark of the top quark mass\footnote{We explain this choice in the context 
of QCD corrections below.}:
\begin{equation}
\Delta^{(K)}_{\rm Box}C_1^{VLL}(\mu_t)=\Delta_1(\mu_t,K)+\Delta_2(\mu_t,K)+\Delta_3(\mu_t,K)\,,
\end{equation}
where
\begin{align}
\label{VLLK1}
&\Delta_1(\mu_t,K)=(c_{d_{L1}}\,c_{d_{L2}})^2\,\sum_{i,j=1,2,3}\lambda_i(K)\,\lambda_j(K)\,
                            c^2_{u_{Li}}\,c^2_{u_{Lj}}\,       F(x_i,x_j)\,,\\
\label{VLLK2}
&\Delta_2(\mu_t,K)=(c_{d_{L1}}\,c_{d_{L2}})^2\,\sum_{i,j=1,2,3}\lambda_i(K)\,\lambda_j(K)\,
                      s^2_{u_{Li}}\,s^2_{u_{Lj}}\,  F(x^\prime_i,x^\prime_j)\,,\\
\label{VLLK3}
&\Delta_3(\mu_t,K)=(c_{d_{L1}}\,c_{d_{L2}})^2\sum_{i,j=1,2,3}\lambda_i(K)\,\lambda_j(K)\,
                  	\left[
			c^2_{u_{Li}}\,s^2_{u_{Lj}}\,  F(x_i,x^\prime_j)+
                      	s^2_{u_{Li}}\,c^2_{u_{Lj}}\,  F(x^\prime_i,x_j)
			\right]\,.
\end{align}
The arguments of the box-functions $F$ are
\begin{equation}
x_i=\left(\frac{m_{u^i}}{M_{W}}\right)^2\,, \qquad 
x^\prime_j=\left(\frac{m_{u^{\prime j}}}{M_{W}}\right)^2\,,
\end{equation}
where both $i$ and $j$ run over $1,2,3$. The loop-function $F(x_i,x_j)$ is 
\begin{equation}
F(x_i,x_j)  =  \dfrac{1}{4}\Big[ \left(4+x_i\,x_j\right)\,I_2\left(x_i,\,x_j\right) - 8\, x_i\, x_j\, I_1\left(x_i,\, x_j\right)\Big]
\end{equation}
with
\begin{equation}
\begin{aligned}
I_1(x_i,\,x_j) &  = & \dfrac{1}{(1-x_i)(1-x_j)} + \left[\frac{x_i\, \ln(x_i)}{(1-x_i)^2 
(x_i-x_j)} + (i\leftrightarrow j)\right]\,,\\
I_2(x_i,\,x_j) & = & \dfrac{1}{(1-x_i)(1-x_j)} + \left[ \frac{x_i^2\, \ln(x_i)}{(1-x_i)^2 
(x_i-x_j)} + (i\leftrightarrow j)\right].
\end{aligned}
\end{equation}
For the $B_{q}^0-\bar B_{q}^0$ mixing we have to replace $K$ by 
$B_q$ and $c_{d_{L1}}\,c_{d_{L2}}$ by $c_{d_{L1}}\,c_{d_{L3}}$ ($c_{d_{L2}}\,c_{d_{L3}}$) in the case of $q=d$ ($q=s$).
There are no contributions to other coefficients from box-diagrams.

\subsection{Wilson Coefficients from Tree-Diagrams}

Calculating the tree-level diagrams in Fig.~\ref{fig:deltaF2tree}
with the exchange of neutral gauge boson mass-eigenstates $\hat{A}^m$ ($m=1,\ldots,24$) we find
the following contributions to the Wilson coefficient at the high scale $\mu_H$, which is of
the order of the mass of the corresponding neutral gauge boson: for the $K$ system we have
\begin{align}
\label{AVLL}
&\Delta^{(K)}_AC_1^{VLL}(\mu_H)=\frac{4\pi^2}{G_F^2M_W^2}\sum_{m=1}^{24}\frac{1}{2\, \hat{M}^2_{A^m}}
\left[\left(\hat \cG^d_L\right)_{ds,m}\right]^2\\
\label{AVRR}
&\Delta^{(K)}_AC_1^{VRR}(\mu_H)=\frac{4\pi^2}{G_F^2M_W^2}\sum_{m=1}^{24}\frac{1}{2\, \hat{M}^2_{A^m}}
\left[\left(\hat \cG^d_R\right)_{ds,m}\right]^2\\
\label{ALR}
&\Delta^{(K)}_AC_1^{LR}(\mu_H)=\frac{4\pi^2}{G_F^2M_W^2}\sum_{m=1}^{24}\frac{1}{2\, \hat{M}^2_{A^m}}
\left[2\,\left(\hat \cG^d_L\right)_{ds,m}\, \left(\hat \cG^d_R\right)_{ds,m} \right]
\end{align}
where the indices $d$ and $s$ stand for the external quarks $d$ and $s$, while the index $m$
refers to the $\hat{A}^m$ gauge boson mass-eigenstate. The corresponding expressions for the
$B_d$ ($B_s$) system are easily derived from the previous ones by substituting $ds$ with $db$
($sb$) in the indices of the couplings. The explicit expression for the couplings
$\left(\hat \cG^d_{L,R}\right)_{ij,m}$ are given  in App.~\ref{app:coup2flavour}.

\subsection{Properties}\label{sec:deltaF2properties}

We note a few properties:
\begin{itemize}
\item[-]
Focussing on eqs.~(\ref{VLLK1})--(\ref{VLLK3}) and the corresponding expressions in the $B_{d,s}$ systems,
for a fixed $\lambda_i\lambda_j$, we have in the box-diagram contributions
the combination
\begin{equation}
\cF_{ij}\equiv c^2_{u_{Li}}c^2_{u_{Lj}} F(x_i,x_j)+s^2_{u_{Li}}s^2_{u_{Lj}}  F(x^\prime_i,x^\prime_j)+
 c^2_{u_{Li}}s^2_{u_{Lj}}  F(x_i,x^\prime_j)+ s^2_{u_{Li}}c^2_{u_{Lj}}  
F(x^\prime_i,x_j).
\label{CalliF}
\end{equation}
If all fermion masses were degenerate, this combination would be 
independent of $i,j$ and the unitarity of the matrix $V$ would
assure the vanishing of FCNC currents. This is precisely what one expects 
from the GIM mechanism.
\item[-]
It is possible to arrange the function $\cF$ in order to match with the usual notation: for the $K$ system we write
\begin{equation}
\label{S0}
\begin{aligned}
S_0(x_t)&\longrightarrow S_t^{(K)}\equiv(c_{d_{L1}}\,c_{d_{L2}})^2 \left(\cF_{33}+\cF_{11}-2\cF_{13}\right)\,,\\
S_0(x_c)&\longrightarrow S_c^{(K)}\equiv(c_{d_{L1}}\,c_{d_{L2}})^2 \left(\cF_{22}+\cF_{11}-2\cF_{12}\right)\,,\\
S_0(x_c,x_t)&\longrightarrow S_{ct}^{(K)}\equiv(c_{d_{L1}}\,c_{d_{L2}})^2 \left(\cF_{23}+\cF_{11}-\cF_{13}-\cF_{12}\right)\,.
\end{aligned}
\end{equation}
For the $B_q$ systems we define similar functions $S_i^{(B_q)}$ that can be simply derived from the previous ones by substituting 
$c_{d_{L1}}\,c_{d_{L2}}$ with $c_{d_{L1}}\,c_{d_{L3}}$ ($c_{d_{L2}}\,c_{d_{L3}}$) in the case of $q=d$ ($q=s$). In particular the combination of the $\cF_{ij}$ factors are universal. In order to recover the $S_0$ functions from the $S^{(M)}_i$ expressions it is necessary to take the limit in which all the cosines are equal to $1$ and all the sines are zero.
\item[-]
The appearance of $c_i$ and $s_j$ factors introduces in general new flavour 
dependence, implying violation of certain MFV relations even in 
the absence of new CP-violating phases.
\item[-]
There are no purely new CP-violating phases in this model, but the CP-odd phase of the CKM matrix induces
sizeable  new effects through new contributions to the mixing induced CP-asymmetries $S_{\psi K_S}$ and $S_{\psi\phi}$, in the 
$B_d^0-\bar B_d$ and the $B_s^0-\bar B_s^0$ systems, respectively. Moreover, similarly to the mass
differences $\Delta M_{B_{d,s}}$, new flavour-violating contributions affect the parameter $\varepsilon_K$
and  there are  correlations between the new physics contributions
to all these observables as we shall see below.
\item[-]
The heavy flavour gauge bosons show flavour-violating couplings that can be
strongly hierarchical: looking at the largest values of these couplings we find
\begin{equation}
\left(\hat \cG^d_{L,R}\right)_{sb}\gg\left(\hat \cG^d_{L,R}\right)_{db}\gg\left(\hat \cG^d_{L,R}\right)_{ds}\,,\quad\quad
\left(\hat \cG^u_{L,R}\right)_{ct}\gg\left(\hat \cG^u_{L,R}\right)_{ut}\gg\left(\hat \cG^u_{L,R}\right)_{uc}\,.
\end{equation}
An example is presented in App.~\ref{app:coup2flavour} for the lightest gauge boson. This hierarchy is due to both the mixings among SM and exotic quarks and the sequential breaking of the flavour symmetry encoded in the flavon VEVs, as seen from Eqs.~(\ref{FGBmasses}), (\ref{DefGunhat}) and (\ref{DefGhat}).
\end{itemize}

\subsection{QCD Corrections and Hadronic Matrix Elements}

The complete analysis requires the inclusion of the renormalisation 
group QCD evolution from the high scales, at which the initial 
effective Hamiltonians given above are constructed, down to low energy 
scales, at which the hadronic matrix elements are evaluated by lattice 
methods. A complication arises in the model in question as several rather 
different high scales are involved, such as the masses of the $W$-boson $M_{W}$,
the masses of the neutral gauge bosons $\hat{M}_{A^m}$ and
the masses of heavy quarks $m_{q^{\prime i}}$.

Before accounting for this problem we recall a very efficient method 
for the inclusion of all these QCD effects in the presence of a single
high scale, which we denote by $\mu_H$. Instead of evaluating the hadronic matrix
elements at the low-energy scale, we can evaluate them at $\mu_H$, corresponding to
the scale at which heavy particles are integrated out.  
The amplitude for $M-\bar{M}$ mixing ($M= K^0, B^0_d,B^0_s$) at the scale $\mu_H$ 
is then simply given by
\begin{equation}
\label{amp6}
\cA(M\to \bar{M})=\dfrac{G_F^2\,M^2_{W}}{4\pi^2}\sum_{i,a} C^a_i(\mu_H)\langle \bar{M} |Q^a_i(\mu_H)|M\rangle\,,
\end{equation}
where the sum runs over all the operators listed in eq.~(\ref{normalM}). The 
matrix element for $M-\bar{M}$ mixing is given by
\begin{equation}
\label{eq:matrix}
\langle \bar M|Q_i^a(\mu_H)|M\rangle = \dfrac{2}{3}\,m_{M}^2\, F_{M}^2\, P_i^a(M),
\end{equation}
where the coefficients $P_i^a(M)$ collect compactly all RG effects from scales below $\mu_H$ as well as
hadronic matrix elements obtained by lattice methods at low energy scales. Analytic formulae for all these coefficients, $P_i^a(B_q)$ and $P_i^a(K)$, are given in Ref.~\cite{Buras:2001ra}, while the corresponding numerical values will be given below for some interesting values of $\mu_H$. 

The question then is how to generalise this method to the case at 
hand which involves several rather different high scales. 
There are three  types 
of contributions for which the relevant high energy scales attributed to 
the coefficients quoted above will differ from each other:
\begin{enumerate} 
\item
 The SM box-diagrams involving $W$-bosons and the SM quarks. Here the scale 
 is chosen to be $\mu_t=\ord(m_t)$.
\item
Tree-level diagrams mediated by neutral heavy gauge bosons, $\hat A^m$. Since we are
taking into consideration the contributions from all such gauge bosons, we shall take
as the initial scale for the RG evolution in each case exactly the mass of the
involved gauge boson.
\item
The only problematic case at first sight are the contributions from box-diagrams 
that involve simultaneously heavy and light particles.
Here the correct procedure would be to integrate out first
the heavy fermions and construct an effective field theory not involving 
them as dynamical degrees of freedom.
However, as the only relevant contribution comes from the lightest 
exotic fermion\footnote{This is strictly true only for the $B_q$ systems, because in the $K$ system due to the CKM suppressions, the contribution from $c'$ may be non-negligible, as accounted for in our numerical analysis. Still, the most relevant contribution comes from $t'$.}, that is $t^\prime$, whose mass is relatively close to $m_t$, we can also here set the matching scale to be $\mu_t$.
As the 
 dominant effects from RG evolution, included here, come from scales 
 below $M_{W}$, this procedure should sufficiently well approximate 
 the exact one.
\end{enumerate}

Having the initial conditions for Wilson coefficients
at a given high scale $\mu_H$ and provided also the corresponding 
hadronic matrix elements at this scale are known, we can calculate 
the relevant $M-\bar{M}$ amplitude by means of eq.~(\ref{amp6}).
As seen in eq.~(\ref{eq:matrix})
these matrix elements 
are directly given in terms of the parameters $P_i^a(K)$, $P_i^a(B_d)$ 
and $P_i^a(B_s)$ for which explicit expressions in terms of RG 
QCD factors and the non-perturbative parameters $B_i^a(\mu_L)$ are given in 
eqs.~(7.28)--(7.34) of Ref.~\cite{Buras:2001ra}: the $\mu_L$ denotes the low energy 
scale and it takes the value $2\gev$ ($4.6\gev$) for the $K$ system ($B_q$ systems).

The $B_i^a(\mu_L)$ parameters are subject to considerable uncertainties. 
Exception are the $B_1^{VLL}$ parameters for which a significant progress 
has been made in the recent years by lattice simulations. 
In the SM analysis, the RG invariant parameters $\hat B_1^{VLL}$ are usually considered and denoted by $\hat B_K$ and $\hat B_{B_q}$. We report their 
values  in tab.~\ref{tab:input}. For completeness we recall the values of 
 $B_1^{VLL}$ that we extracted
from the most recent lattice simulations:
\begin{equation}
\begin{aligned}
&B_1^{VLL} =  0.515(14)\,,\qquad\qquad
&&\text{for $K$ system}\\
&B_1^{VLL} = 0.825(72)\,,\qquad\qquad
&&\text{for $B_d$ system}\\
&B_1^{VLL} = 0.871(39)\,,\qquad\qquad
&&\text{for $B_s$ system}\,.
\end{aligned}
\end{equation}
As  these parameters are the same for VRR contributions we will combine them together with
the VLL contributions in the final formula at the end of this section.  

Neglecting the unknown $\ord(\alpha_s)$ contributions to Wilson coefficients of the remaining 
operators at the high energy scale, our NLO RG analysis 
involves only the values of the coefficients $P_1^{LR}(K)$, $P_1^{LR}(B_d)$ and $P_1^{LR}(B_s)$ calculated at $\mu_H$. We are not considering the intermediate thresholds of exotic quarks, since the smallness of $\alpha_s$ and the absence of the flavour dependence in the LO anomalous dimensions of the contributing operators render the corresponding effects negligible. To obtain these values we need only the values of $B_1^{LR}$ and $B_2^{LR}$, that we report below in the NDR scheme\footnote{These values can be found in Refs.~\cite{Babich:2006bh,Becirevic:2001xt}, where $B_1^{LR}$ ($B_2^{LR}$) is called $B_5$ ($B_4$).} \cite{Babich:2006bh,Becirevic:2001xt}:
\begin{equation}
\begin{aligned}
&B_1^{LR} =  0.562(39)(46)\,,\qquad
&&B_2^{LR} =  0.810(41)(31)\,,\qquad
&&\text{for $K$ system}\\
&B_1^{LR} = 1.72(4)({}^{+20}_{-6})\,,\qquad
&&B_2^{LR} = 1.15(3)({}^{+5}_{-7})\,,\qquad
&&\text{for $B_d$ system}\\
&B_1^{LR} = 1.75(3)({}^{+21}_{-6})\,,\qquad
&&B_2^{LR} = 1.16(2)({}^{+5}_{-7})\,,\qquad
&&\text{for $B_s$ system}\,.
\end{aligned}
\end{equation}
In tab.~\ref{tab:PiFactors}, we show the resulting $P_i$ factors for some relevant values of $\mu_H$.

\begin{table}[ht]
\begin{center}
\begin{tabular}{|l||c|c|c|c|}
\hline
&&&&\\[-3mm]
$~\qquad\mu_H$ 				& $500\gev$ & $1\tev$ & $3\tev$ 	& $10\tev$ \\[2mm]
\hline
&&&&\\[-3mm]
$P_1^{VLL}(\mu_H,\,K)$ 		& 0.392 		& 0.384 	&  0.373 	& 0.363 \\[2mm]
$P_1^{LR}(\mu_H,\,K)$ 			& -35.7  		& -39.3 		& -45.0 		& -51.4 \\[2mm]
$P_1^{VLL}(\mu_H,\,B_d)$ 	& 0.675 		& 0.662 	& 0.643 	& 0.624 \\[2mm]
$P_1^{LR}(\mu_H,\,B_d)$ 		& -2.76  		& -2.97 		& -3.31 		& -3.69 \\[2mm]
$P_1^{VLL}(\mu_H,\,B_s)$ 	& 0.713 		& 0.698 	& 0.678 	& 0.659 \\[2mm]
$P_1^{LR}(\mu_H,\,B_s)$ 		& -2.76 			& -2.97 		& -3.31 		& -3.69\\[2mm]
\hline
\end{tabular}
\end{center}
\caption{\it Central values of $P_i$ factors for \mbox{$\mu_H=\{0.5,\,1,\,3,\,10\}\tev$.}}
\label{tab:PiFactors}
\end{table}

Notice that the LR operators, arising from integrating out the heavy flavour gauge bosons, are strongly
enhanced by the RG QCD running as can be deduced from the values of the $P_1^{LR}$ factors.
A priori, such contributions could be very important.

\subsection[\texorpdfstring{Final Formulae for $\Delta F=2$ Observables}
                           {Final Formulae for Delta F=2 Observables}]{
\mathversion{bold}Final Formulae for $\Delta F=2$ Observables\mathversion{normal}
}

We collect here the formulae we shall use in our numerical analysis. The mixing amplitude $M^i_{12}$ $(i=K,d,s)$ is related to the relevant effective Hamiltonian through
\begin{equation}
2\,m_K\left(M_{12}^K\right)^\ast=
\langle\bar K^0|\Heff^{\Delta S=2}|K^0\rangle\,,\qquad\qquad
2\,m_{B_q}\left(M_{12}^q\right)^\ast=
\langle\bar B_q^0|\Heff^{\Delta B=2}|B_q^0\rangle
\label{MixingAmplitudeHamiltonian}
\end{equation}
with $q=d,s$.
The $K_L-K_S$ mass difference and the CP-violating parameter $\varepsilon_K$ are then given by
\begin{equation}
\Delta M_K=2\RE\left(M_{12}^K\right)\,,\qquad\qquad
\varepsilon_K=\dfrac{\kappa_\eps\, e^{i\,\varphi_\eps}}{\sqrt{2}(\Delta M_K)_\text{exp}}\IM\left(M_{12}^K\right)\,,
\label{DeltaMKepsilonK}
\end{equation}
where $\varphi_\eps = (43.51\pm0.05)^\circ$ and $\kappa_\eps=0.923\pm0.006$ takes into account that $\varphi_\eps\ne \pi/4$ and includes long distance effect in $\IM \Gamma_{12}$ \cite{Buras:2008nn} \footnote{This value has been confirmed by lattice and presented with a smaller error in Ref.~\cite{Blum:2011ng}.} and $\IM M_{12}$ \cite{Buras:2010pza}. The mixing amplitude entering the previous expressions can be decomposed into two parts, one containing the $LL$ and $RR$ contributions and the second only the $LR$ ones: 
\begin{equation}
M^K_{12}=(M^K_{12})_1+(M^K_{12})_2,
\end{equation}
where
\begin{equation}
\begin{aligned}
&\begin{split}
(M_{12}^K)_1=\dfrac{G_F^2\,M_W^2}{12\pi^2}\,F_K^2\,m_K\Big[&\hat{B}_K\,\eta_1\,\lambda^2_2(K)\,S^{(K)}_c+\hat{B}_K\,\eta_2\,\lambda^2_3(K)\,S^{(K)}_t+\\
&+2\,\hat{B}_K\,\eta_3\,\lambda_2(K)\,\lambda_3(K)\,S_{ct}^{(K)}+\\
&+P_1^{VLL}(\mu_H,K)\,\left(\Delta^{(K)}_A\,C_1^{VLL}(\mu_H)+\Delta^{(K)}_A\,C^{VRR}(\mu_H)\right)\,
\Big]^*\,,
\end{split}\\
&(M_{12}^K)_2=\dfrac{G_F^2\,M_W^2}{12\pi^2}\,F_K^2\,m_K\,P_1^{LR}(\mu_H,K)\,\Delta^{(K)}_A\,C_1^{LR*}(\mu_H)\,.
\end{aligned}
\end{equation}

Analogously, for the the $B_{d,s}^0-\bar B_{d,s}^0$ systems the two parts of the mixing amplitude are given by:
\begin{equation}
\begin{aligned}
&\begin{split}
(M_{12}^{q})_1=\dfrac{G_F^2\,M^2_W}{12\pi^2}\,F^2_{B_q}\,m_{B_q}\,
\Big[&\eta_B\,\hat{B}_{B_q}\,\lambda^2_3(B_q)\,\,S_t^{(B_q)}+\\
&+P_1^{VLL}(\mu_H,B_q)\,\left(\Delta^{(B_q)}_A C_1^{VLL}(\mu_H)+\Delta^{(B_q)}_A C^{VRR}(\mu_H)\right)\Big]^*\,,
\end{split}\\
&(M_{12}^{q})_2=\frac{G_F^2\,M^2_W}{12\pi^2}\,F^2_{B_q}\,m_{B_q} \,
P_1^{LR}(\mu_H,B_q)\,\Delta^{(B_q)}_A C_1^{LR*}(\mu_H)\,.
\end{aligned}
\end{equation}
Here $\eta_{1,2,3,B}$ are known SM QCD corrections given in tab.~\ref{tab:input} and
$P_i^a(\mu_H,M)$ describe the QCD evolution from $\mu_H$ down to $\mu_L$ for the considered system. 
For the $B_q$ systems, it is useful to rearrange the definition of the mixing amplitude $M_{12}^q$ as follows \cite{Bona:2005eu}
\begin{equation}
M_{12}^q=\left(M_{12}^q\right)_\text{SM}C_{B_q}e^{2\,i\,\varphi_{B_q}}\,,
\label{NewMixingAmplitudeBds}
\end{equation}
where $C_{B_{d,s}}$ and $\varphi_{B_{d,s}}$ account for deviations from the SM contributions. Therefore, the mass differences turn out to be 
\begin{equation}
\Delta M_{B_q}=2\left|M_{12}^q\right|=(\Delta M_{B_q})_\text{SM}C_{B_q}\qquad (q=d,s)\,,
\label{DeltaMB}
\end{equation}
where
\begin{equation}
\bigl(M_{12}^d\bigr)_\text{SM}= \bigl|\bigl(M_{12}^d\bigr)_\text{SM}\bigr|e^{2\,i\,\beta}\,,\qquad\qquad
\bigl(M_{12}^s\bigr)_\text{SM}= \bigl|\bigl(M_{12}^s\bigr)_\text{SM}\bigr|e^{2\,i\,\beta_s}\,.
\label{MB12SM}
\end{equation}
Here the phases $\beta\approx 22^\circ$ and $\beta_s\simeq -1^\circ$ are defined through
\begin{equation}
V^{SM}_{td}=|V_{td}^{SM}|e^{-i\beta}
\qquad\qquad\textrm{and}\qquad\qquad 
V^{SM}_{ts}=-|V^{SM}_{ts}|e^{-i\beta_s}\,.
\label{eq:3.40}
\end{equation}
The coefficients of $\sin(\Delta M_{B_d}\, t)$ and $\sin(\Delta M_{B_s}\, t)$ in the time dependent asymmetries in $B_d^0\to\psi K_S$ and $B_s^0\to\psi\phi$ 
are then given, respectively, by:
\begin{equation}
S_{\psi K_S} =\sin(2\beta+2\varphi_{B_d})\,,\qquad\qquad
S_{\psi\phi} = \sin(2|\beta_s|-2\varphi_{B_s})\,.
\label{Sobservables}
\end{equation}
Notice that in the presence of non-vanishing $\varphi_{B_d}$ and $\varphi_{B_s}$ these two asymmetries do not measure $\beta$ and $\beta_s$ but $(\beta+\varphi_{B_d})$ and $(|\beta_s|-\varphi_{B_s})$, respectively.

\mathversion{bold}
\subsection{\texorpdfstring
{The Ratio $\Delta M_{B_d}/\Delta {M_{B_s}}$ and the $B^+\to\tau^+\nu$ Decay}
{The Ratio Delta MBd/Delta MBs and the B^+ -> tau+ nu Decay}}
\mathversion{normal}

The expressions for the mass-differences recovered in the previous section are affected by
large uncertainties, driven by the decay constants $F_{B_{d,s}}$. To soften the
dependence of our analysis on these theoretical errors, we consider the ratio among
$\Delta M_{B_d}$ and $\Delta M_{B_s}$, that we call $R_{\Delta M_B}$, and the ratio among
the branching ratio of the $B^+\to\tau^+\nu$ decay and $\Delta M_{B_d}$, that we name
$R_{BR/\Delta M}$. 

Indeed, when considering $R_{\Delta M_B}$, we notice that the SM theoretical errors are
encoded into the parameter $\xi=1.237\pm0.032$, that is much less affected by uncertainties
with respect to the mass differences. When considering the NP effects, we obtain
\begin{equation}
R_{\Delta M_B}=\dfrac{(\Delta M_{B_d})_\text{SM}}{(\Delta M_{B_s})_\text{SM}}\dfrac{C_{B_d}}{C_{B_s}}\,.
\end{equation}

On the other hand, in the SM, the $B^+\to\tau^+\nu$ decay occurs at the tree-level through the
exchange of the $W$-boson. Therefore, the expression for its branching ratio is only slightly modified
in our model: 
\begin{equation}
BR(B^+\to\tau^+\nu)=\dfrac{G_F^2\,m_{B^+}\,m_\tau^2}{8\pi}\left(1-\dfrac{m_\tau^2}{m^2_{B^+}}\right)^2\,F^2_{B^+}\,|c_{u_{L1}} \,V_{ub}\,c_{d_{L3}}|^2\,\tau_{B^+}\,,
\end{equation}
where the NP effects are represented by the cosines. Notice that heavy flavour gauge boson
contributions could contribute only at the loop-level  and can be safely neglected,
since they compete with a tree-level process. Furthermore, in the previous expression we
have assumed the SM couplings for the leptons to the $W$-boson: even if we are not considering
the lepton sector in our analysis it is reasonable to assume that any NP modification can be
safely negligible, as these couplings are strongly constrained by the SM electroweak analysis.

In the ratio $R_{BR/\Delta M}$, the dependence on $F_{B_d}$, which is indeed the main source
of the theoretical error on $\Delta M_{B_d}$, is cancelled \cite{Ikado:2006un,Isidori:2006pk}:
\begin{equation}
R_{BR/\Delta M}=\dfrac{3\,\pi\,\tau_{B^+}}{4\,\eta_B\,\hat B_{B_d}\,S_0(x_t)}\dfrac{c^2_{u_{L1}}\,c^2_{d_{L3}}}{C_{B_d}}\dfrac{m^2_\tau}{M_W^2}\dfrac{\left|V_{ub}\right|^2}{\left|V^*_{tb}\,V_{td}\right|^2}\left(1-\dfrac{m_\tau^2}{m_{B_d}^2}\right)^2\,,
\end{equation} 
where the second fraction contains all NP contributions and we took $m_{B^+}\approx m_{B_d}$, well justified considering the errors in the other quantities. The SM prediction of this observable should be compared with the data 
\begin{equation}
R_{BR/\Delta M}=(3.25\pm0.67)\times 10^{-4}\ps\,.
\end{equation}
See tab.~\ref{tab:predictionsexperiment} for the SM prediction.

%
%

\mathversion{bold}
\section[\mathversion{bold} The $b$ semileptonic CP-asymmetry \mathversion{normal}]{The $b$ semileptonic CP-asymmetry}
\mathversion{normal}
\label{sec:ASL}

In the $B_q$ systems, apart from $\Delta M_{B_q}$, $S_{\psi K_S} $ and $S_{\psi\phi}$, a third
quantity providing information on the meson mixings is the $b$ semileptonic CP-asymmetry
$A^b_{sl}$ \cite{Abazov:2011yk,Lenz:2011ww}:
\begin{equation}
A^b_{sl}=(0.594\pm0.022)\,a^d_{sl}+(0.406\pm0.022)\,a^s_{sl}\,,
\end{equation}
where
\begin{equation}
\begin{aligned}
&a^d_{sl}=\left|\dfrac{\left(\Gamma_{12}^d\right)_{SM}}{\left(M_{12}^d\right)_{SM}}\right|\sin\phi_d=(5.4\pm1.0)\times10^{-3}\,\sin\phi_d\,,\\
&a^s_{sl}=\left|\dfrac{\left(\Gamma_{12}^s\right)_{SM}}{\left(M_{12}^s\right)_{SM}}\right|\sin\phi_s=(5.0\pm1.1)\times10^{-3}\,\sin\phi_s\,,
\end{aligned}
\end{equation}
with
\begin{equation}
\begin{aligned}
&\phi_d=\arg\Big(-\left(M_{12}^d\right)_{SM}/\left(\Gamma_{12}^d\right)_{SM}\Big)=-4.3^\circ\pm 1.4^\circ\,,\\
&\phi_s=\arg\Big(-\left(M_{12}^s\right)_{SM}/\left(\Gamma_{12}^s\right)_{SM}\Big)=0.22^\circ\pm 0.06^\circ\,.
\end{aligned}
\end{equation}

In the presence of NP, these expressions are modified. Since we have already discussed the NP effects on $M_{12}^q$ in the previous sections, we focus now only on $\Gamma_{12}^q$. It is useful to adopt a notation for $\Gamma_{12}^q$ similar to the one in eq.~(\ref{NewMixingAmplitudeBds}) for $M_{12}^q$:
\begin{equation}
\Gamma_{12}^q=(\Gamma_{12}^q)_\text{SM}\,\tilde C_{B_q}\,e^{-2\,i\,\tilde\varphi_{B_q}}\,,
\label{NotationGamma12withNP}
\end{equation}
where $\tilde C_{B_q}$ is a real parameter. With such a notation we get,
\begin{equation}
a^q_{sl}=\left|\dfrac{\left(\Gamma_{12}^d\right)_{SM}}{\left(M_{12}^d\right)_{SM}}\right|\dfrac{\tilde C_{B_q}}{C_{B_q}}\sin\left(\phi_d+2\varphi_{B_q}+2\tilde \varphi_{B_q}\right)\,.
\end{equation}
Notice, that in the MGF context we are considering, the phase $\tilde\varphi_{B_q}$ is vanishing, while $\tilde C_{B_q}$ is mainly given by $c^2_{u_{L2}}\,c_{d_{Lb}}\,c_{d_{Lq}}\approx1$. As a result the only NP modifications are provided by the NP contributions on $M^q_{12}$.

%
%

\mathversion{bold}
\section[\texorpdfstring{
         \mathversion{bold}The $\bar B\to X_s \gamma$ Decay\mathversion{normal}}
	 {The B -> Xs gamma Decay}]{The $\bar B \to X_s \gamma$ Decay}
\mathversion{normal}
\label{sec:BSG}

\subsection{Effective Hamiltonian}
The decay $\bar B\to X_s\gamma$ is mediated by the photonic dipole operators
$Q_{7\gamma}$ and $Q_{7\gamma}^{\prime}$  and through mixing also by the
gluonic dipole operators $Q_{8G}$ and $Q_{8G}^{\prime}$. In our conventions they read
\begin{equation}
\label{O6B}
\begin{aligned}
Q_{7\gamma}  &=  \dfrac{e}{16\pi^2}\, m_b\, \bar{s}_\alpha\, \sigma^{\mu\nu}\, P_R\, b_\alpha\, F_{\mu\nu}\,,\\[2mm]          
Q_{8G}     &=  \dfrac{g_s}{16\pi^2}\, m_b\, \bar{s}_\alpha\, \sigma^{\mu\nu}\, P_R\, T^a_{\alpha\beta}\, b_\beta\, G^a_{\mu\nu}  
\end{aligned}
\end{equation}
and the corresponding primed dipole operators are obtained by substituting 
$P_R$ with $P_L$.

The effective Hamiltonian for $b\to s\gamma$ at a scale $\mu$ in the SM
normalisation and considering only the dipole operators reads
\begin{equation}
\label{HeffW_at_mu}
\begin{split}
{\cal H}_{\text{eff}}^{b\to s\gamma} = 
- \dfrac{4 G_{\rm F}}{\sqrt{2}} V_{ts}^* V_{tb} \Big[&
\Delta C_{7\gamma}(\mu) Q_{7\gamma} + \Delta  C_{8G}(\mu) Q_{8G}+\\
&+\Delta C^{\prime}_{7\gamma}(\mu) Q^{\prime}_{7\gamma} + \Delta  C^{\prime}_{8G}(\mu) Q^{\prime}_{8G}
\Big]\,.
\end{split}
\end{equation}
We have kept the contributions of the primed dipole operators
$Q_{7\gamma}^{\prime}$ and $Q_{8G}^{\prime}$ even though their Wilson coefficients 
are suppressed by $m_s/m_b$ with respect to the unprimed Wilson coefficients.
However, the mixing of neutral current-current operators into
$Q_{7\gamma}^{\prime}$ and $Q_{8G}^{\prime}$ can affect $\Delta
C^{\prime}_{7\gamma}(\mu_b)$ as shown in Ref.~\cite{Buras:2011zb}.

Similarly to the Hamiltonian for the $\Delta F=2$ transitions, the Wilson coefficients
in the Hamiltonian can be separated into two parts:
\begin{itemize}
  \item[-] The SM-like contribution from diagrams with $W$-bosons with modified couplings
    to both SM and exotic quarks of charge $+2/3$, denoted below by $u$ and $u'$,
    respectively:
  \begin{center}
  \includegraphics{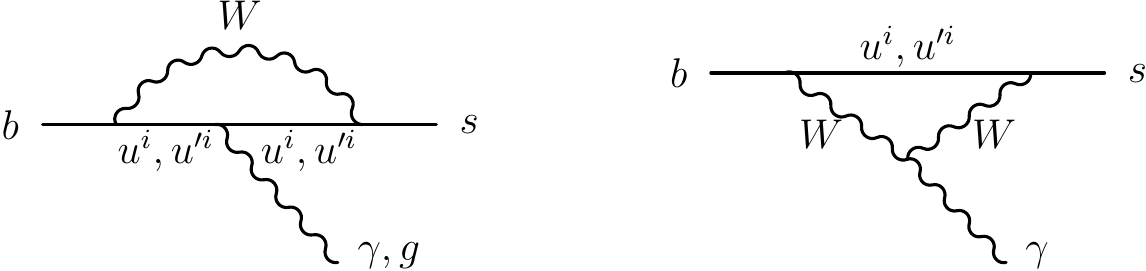}
  \end{center}

\item[-] The contribution of heavy neutral gauge bosons exchanges with virtual SM and exotic quarks of
  charge $-1/3$, denoted below by $d$ and $d'$, respectively:
  \begin{center}
  \includegraphics{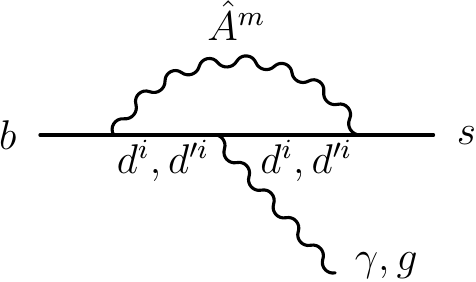}
  \end{center}
\end{itemize}

The first contribution has already been considered in Ref.~\cite{Grinstein:2010ve}, while the second, the impact of the heavy
neutral gauge bosons on $b\rightarrow s\gamma$, has been recently pointed out in Ref.~\cite{Buras:2011zb}. In particular
it has been found that the QCD renormalisation group effects in the neutral gauge boson contributions can strongly affect the branching ratio of $\bar B\to X_s\gamma$ and cannot be neglected a priori.

\mathversion{bold}
\subsection{Contributions of $W$-exchanges}
\label{subsec:HeffBsgammaW}
\mathversion{normal}

For the $W$-exchange the matching is performed at the EW scale, $\mu_W$. The Wilson
coefficients are the sum of $t$ and $t^\prime$ contribution, since
$c^{\prime}$ and $u^{\prime}$ contributions are suppressed by their small
couplings to $b$ and $s$ quarks. Hence, the Wilson coefficients of $Q_{7\gamma}$ and $Q_{8G}$
are
\begin{align}
\Delta_W C_{7\gamma}(\mu_W)&=
c_{d_{L2}}\,c_{d_{L3}}
\left(  c^2_{u_{L3}} C_{7\gamma}^{SM}(x_t) + s^2_{u_{L3}} C_{7\gamma}^{SM}(x_t^\prime)\right)\,,\\
\Delta_W C_{8G}(\mu_W)&=
c_{d_{L2}}\,c_{d_{L3}}
\left(  c^2_{u_{L3}} C_{8G}^{SM}(x_t) + s^2_{u_{L3}} C_{8G}^{SM}(x_t^\prime)\right)\,,
\end{align}
with 
\begin{align}
C^{SM}_{7\gamma} (x) &= \dfrac{3 x^3-2 x^2}{4(x-1)^4}\ln x - \frac{8 x^3 + 5 x^2 - 7 x}{24(x-1)^3}\,,\\
C^{SM}_{8G}(x) &= \dfrac{-3 x^2}{4(x-1)^4}\ln x - \frac{x^3 - 5 x^2 - 2 x}{8(x-1)^3}
\end{align}
being the SM Inami-Lim functions \cite{Inami:1980fz}.

At last we need to evolve $\Delta_W C(\mu_W)$ down to $\mu_b$ to obtain the
contribution of $W$ exchanges to the branching ratio of $\bar B\to X_s\gamma$. The QCD analysis, that involves 
the SM charged current-current operators $Q_1$ and $Q_2$ as well as the
QCD-penguins $Q_3$ to $Q_6$, which mix with $Q_{7\gamma}$ and $Q_{8G}$ below $\mu_t$,
is the same as in the SM and we proceed as in Ref.~\cite{Buras:2011zb}.

\mathversion{bold}
\subsection{\texorpdfstring
          {Contributions of $\hat{A}^m$-exchanges}
	  {Contributions of A-exchanges}}
\label{subsec:HeffBsgammaAH}
\mathversion{normal}

The contribution of a neutral gauge boson to the effective Hamiltonian in 
eq.~(\ref{HeffW_at_mu})
derives from integrating out the mass-eigenstate of the heavy flavour
gauge boson $\hat{A}^m$ at its  mass-scale $\mu_H$. The Wilson coefficients 
$\Delta_A C^{(\prime)}_{7\gamma}(\mu_H)$ and $\Delta_A C^{(\prime)}_{8G}(\mu_H)$ 
have been calculated within a generic framework in Ref.~\cite{Buras:2011zb}.
The results for the special MGF case we are discussing are fixed by the couplings of the
flavour gauge bosons to both SM and exotic fermions. Applying general 
formulae of Ref.~\cite{Buras:2011zb} to the present case we found that these 
contributions are below $1\%$ and can be safely neglected. As discussed in Ref.~\cite{Buras:2011zb}
the reason for such suppression are the See-saw-like couplings of flavour gauge bosons to both SM and exotic fermions and the heavy neutral gauge boson masses.

%
%
\section{Numerical Analysis}
Having at hand the analytic expressions derived in the previous sections, 
we are ready to perform a numerical analysis of the MGF model in question.

The first question we ask is whether the model is able to remove various anomalies in the flavour 
data hinting the presence of NP. Since the number of parameters is much smaller than in
other popular extensions of the SM like
SUSY models, LHT model, RS-scenario and models with left-right symmetry, it is indeed not obvious
that these anomalies can be removed or at least softened. We briefly review the flavour
anomalies as seen from the SM point of view.

\subsection{Anomalies in the Flavour Data}

\boldmath
\subsubsection{\texorpdfstring
 {The $\varepsilon_K-S_{\psi K_S}$ Anomaly}
 {The e_K - SpsiKs Anomaly}}
\unboldmath

It has been pointed out in Refs.~\cite{Lunghi:2008aa,Buras:2008nn,Buras:2009pj,Lenz:2010gu}
that the SM prediction for $\varepsilon_K$ implied by the measured value of $S_{\psi K_S}=\sin 2\beta$,
the ratio $R_{\Delta M_B}$, and the value of $|V_{cb}|$ is too small to agree well with
the experiment. We obtain the SM $\varepsilon_K$ value by taking the experimental value of
$S_{\psi K_S}$, the ultimate $\vcb$ determination, the most recent value of the non-perturbative
parameter $\hat B_K$ \cite{Antonio:2007pb,Aubin:2009jh,Laiho:2009eu,Bae:2010ki,Constantinou:2010qv,Aoki:2010pe}, and by including long-distance effects in ${\rm Im}\Gamma_{12}$ \cite{Buras:2008nn} and
${\rm Im}M_{12}$ \cite{Buras:2010pza} as well as recently calculated NNLO
QCD corrections to $\varepsilon_K$ \cite{Brod:2010mj,Brod:2011ty}.
We find\footnote{The small discrepancy with respect to the value of Ref.~\cite{Brod:2011ty}
comes solely from updated input values.} $|\varepsilon_K|=(1.82\pm0.28)\times 10^{-3}$,
visibly below the experimental value.

On the other hand $\sin 2\beta=0.85\pm 0.05$ from SM  fits of the Unitarity Triangle is significantly
larger than the experimental value. This discrepancy is  to some extent caused by the desire to fit
both $\varepsilon_K$ \cite{Lunghi:2008aa,Buras:2008nn,Buras:2009pj} and $BR(B^+\to\tau^+\nu)$
\cite{Lunghi:2010gv}.

As demonstrated in \cite{Buras:2008nn,Buras:2009pj}, whether the NP is required in $\varepsilon_K$
or  $S_{\psi K_S}$ depends on the values of $\gamma$, $|V_{ub}|$ and $\vcb$. The phase $\gamma$ 
should be measured precisely by LHCb in the coming years while $\vub$ and $\vcb$ should
be precisely determined by Belle II and Super-$B$ provided that also the hadronic
uncertainties will be under a better control. 

\boldmath
\subsubsection{The $|V_{ub}|$-Problem}
\unboldmath
There is a tension between inclusive and exclusive determinations of $|V_{ub}|$. This means
that if we take the unitarity of the CKM matrix as granted and also consider the good agreement of the ratio
$R_{\Delta M_B}$ with the data, the inclusive and exclusive determinations imply different
patterns of NP in CP-violating observables. Indeed one is lead to consider two limiting scenarios:
\begin{description}
\item[{\boldmath Scenario 1: Small $|V_{ub}|.$}]
Here $|V_{ub}|$ is in principle the exclusive determination,
\begin{equation}
|V_{ub}|=(3.38\pm0.36)\times 10^{-3}\,.
\end{equation}
Within the SM, when the $\Delta M_{B_s}/\Delta M_{B_d}$ constraint is taken into account, one finds
$S_{\psi K_S}\approx 0.67$ in agreement with the data, but $\varepsilon_K\approx 1.8\times 10^{-3}$
visibly below the data. As discussed in Refs.~\cite{Buras:2008nn,Buras:2009pj}, a sizeable constructive
NP contribution to $\varepsilon_K$ would not require an increased value of $\sin 2\beta$ relative
to the experimental value of $S_{\psi K_S}$. 
NP of this type would then remove the $\varepsilon_K-S_{\psi K_S}$ anomaly in the presence
of the exclusive value of $\vub$.

\item[{\boldmath Scenario 2: Large $|V_{ub}|$.}]
In this case  $|V_{ub}|$ corresponds to its inclusive determination,
\be
|V_{ub}|=(4.27\pm0.38)\times 10^{-3}\,.
\ee
In this scenario  the SM predicts $\varepsilon_K\approx 2.2\times 10^{-3}$, in agreement with the data, while
$S_{\psi K_S}\approx 0.81$ is significantly above the data. As discussed in Refs.~\cite{Lunghi:2008aa,Buras:2008nn},
a negative NP phase $\varphi_{B_d}$ in $B^0_d-\bar B^0_d$ mixing would solve the
$\varepsilon_K-S_{\psi K_S}$ anomaly in this case (see eq.~(\ref{Sobservables})),
provided such a phase is phenomenologically allowed by other constraints. With a negative $\varphi_{B_d}$,
$\sin 2\beta$ is larger than $S_{\psi K_S}$, implying a higher value on $|\varepsilon_K|$,
in reasonable agreement with data and a better Unitary Triangle fit. 
\end{description}

In both  scenarios, new physics contributions to other observables, such as $\Delta M_{B_{d,s}}$,
are expected and a dedicate analysis is necessary. In fact as we will see 
below the correlations between $\varepsilon_K$ and $\Delta M_{B_{d,s}}$ are powerful tests of the ability of MGF to describe properly all data on 
$\Delta F=2$ observables.

\subsection{Input Parameters and the Parameter Space of the Model}

Before proceeding with our numerical analysis, it is necessary to fix the input
parameters and to define the parameter space of the model.  

\subsubsection{Input Parameters}

\begin{table}[ht|]
\renewcommand{\arraystretch}{1}\setlength{\arraycolsep}{1pt}
\center{\begin{tabular}{|l|l|}
\hline
$G_F = 1.16637(1)\times 10^{-5}\gev^{-2}$\hfill\cite{Nakamura:2010zzi} 	&  $m_{B_d}= 5279.5(3)\mev$\hfill\cite{Nakamura:2010zzi}\\
$M_W = 80.399(23) \gev$\hfill\cite{Nakamura:2010zzi}  							& $m_{B_s} = 5366.3(6)\mev$\hfill\cite{Nakamura:2010zzi}\\
$\sin^2\theta_W = 0.23116(13)$\hfill\cite{Nakamura:2010zzi} 						& $F_{B_d} = 205(12)\mev$\hfill\cite{Laiho:2009eu}\\
$\alpha(M_Z) = 1/127.9$\hfill\cite{Nakamura:2010zzi}							& $F_{B_s} = 250(12)\mev$\hfill\cite{Laiho:2009eu}\\
$\alpha_s(M_Z)= 0.1184(7) $\hfill\cite{Nakamura:2010zzi}						& $\hat B_{B_d} = 1.26(11)$\hfill\cite{Laiho:2009eu}\\\cline{1-1}
$m_u(2\gev)=1.7\div3.1\mev $ 	\hfill\cite{Nakamura:2010zzi}						& $\hat B_{B_s} = 1.33(6)$\hfill\cite{Laiho:2009eu}\\
$m_d(2\gev)=4.1\div5.7\mev$	\hfill\cite{Nakamura:2010zzi}						& $F_{B_d} \sqrt{\hat B_{B_d}} = 233(14)\mev$\hfill\cite{Laiho:2009eu}\\
$m_s(2\gev)=100^{+30}_{-20} \mev$	\hfill\cite{Nakamura:2010zzi}					& $F_{B_s} \sqrt{\hat B_{B_s}} = 288(15)\mev$\hfill\cite{Laiho:2009eu}\\
$m_c(m_c) = (1.279\pm 0.013) \gev$ \hfill\cite{Chetyrkin:2009fv}					& $\xi = 1.237(32)$\hfill\cite{Laiho:2009eu}\\
$m_b(m_b)=4.19^{+0.18}_{-0.06}\gev$\hfill\cite{Nakamura:2010zzi} 					& $\eta_B=0.55(1)$\hfill\cite{Buras:1990fn,Urban:1997gw}\\
$M_t=172.9\pm0.6\pm0.9 \gev$\hfill\cite{Nakamura:2010zzi} 						& $\tau_{B^\pm}=(1641\pm8)\times10^{-3}\ps$\hfill\cite{Nakamura:2010zzi}\\\hline
$m_K= 497.614(24)\mev$	\hfill\cite{Nakamura:2010zzi}							&$|V_{us}|=0.2252(9)$\hfill\cite{Nakamura:2010zzi}\\	
$F_K = 156.0(11)\mev$\hfill\cite{Laiho:2009eu}								&$|V_{cb}|=(40.6\pm1.3)\times 10^{-3}$\hfill\cite{Nakamura:2010zzi}\\
$\hat B_K= 0.737(20)$\hfill\cite{Laiho:2009eu}								&$|V^\text{incl.}_{ub}|=(4.27\pm0.38)\times10^{-3}$\hfill\cite{Nakamura:2010zzi}\\
$\kappa_\epsilon=0.94(2)$\hfill\cite{Buras:2010pza}							&$|V^\text{excl.}_{ub}|=(3.38\pm0.36)\times10^{-3}$\hfill\cite{Nakamura:2010zzi}\\	
$\varphi_\epsilon=(43.51\pm0.05)^\circ$\hfill\cite{Buras:2008nn}					&$\gamma=(73^{+22}_{-25})^\circ$\hfill\cite{Nakamura:2010zzi}\\
$\eta_1=1.87(76)$\hfill\cite{Brod:2011ty}								& \\		
$\eta_2=0.5765(65)$\hfill\cite{Buras:1990fn}								& \\
$\eta_3= 0.496(47)$\hfill\cite{Brod:2010mj}								& \\\hline
\end{tabular}  } 
\caption{Values of experimental and theoretical quantities used throughout our numerical
analysis. Notice that $m_i(m_i)$ are the masses $m_i$ at the scale $m_i$ in the $\ov{MS}$
scheme. $M_t$ is the pole top-quark mass.
\label{tab:input}}
\renewcommand{\arraystretch}{1.0}
\end{table}

In Table~\ref{tab:input} we list the nominal values of the input parameters
that we will use for the numerical analysis, except when otherwise stated.
At this stage it is important to recall the theoretical and the experimental
uncertainties on some relevant parameters and on the observables we shall
study.

Considering the $K^0-\bar K^0$ mixing, remarkable improvements have been made
in the case of the CP-violating parameter $\varepsilon_K$, where the decay
constant $F_K$ is known within $1\%$ accuracy. Moreover the parameter $\hat{B}_K$
is known within $3\%$ accuracy from lattice calculations with dynamical fermions
{\cite{Antonio:2007pb}} and an improved estimate of {\it long distance} contributions to
$\varepsilon_K$ reduced this uncertainty down to $2\%$ \cite{Buras:2008nn,Buras:2010pza}.
The NNLO QCD corrections to $\eta_1$ and $\eta_3$ \cite{Brod:2010mj,Brod:2011ty} allowed
to access the remaining scale uncertainties that amount according to \cite{Brod:2011ty} to
roughly $6\%$, dominantly due to the uncertainty in $\eta_1$.
Including also parametric uncertainties, dominated by the value of
$\vcb$, Brod and Gorbahn estimate
conservatively the present error in $\varepsilon_K$ to amount to roughly $15\%$ \cite{Brod:2011ty}.
The reduction of this total error down to $7\%$  in the coming years appears
to be realistic. Further reduction will require progress both in the evaluation
of long distance contributions and in $\eta_1$.
$\Delta M_K$ is very accurately measured, but is subject to poorly
known long distance contributions.

Regarding the $B_q^0-\bar B_q^0$ mixings, lattice calculations considerably improved
in recent years reducing the uncertainties in 
$F_{B_s}$ \footnote{Recently a remarkably precise value for $F_{B_s}$ was
reported in Ref.~\cite{McNeile:2011ng}: $F_{B_s}=(225\pm4)\mev$. Still, we
shall adopt a conservative approach and  use the value of $F_{B_s}=(250\pm12)\mev$
in our analysis except when explicitly stated.} and $F_{B_d}$ and also in $\sqrt{\hat B_{B_s}}F_{B_s}$ and
$\sqrt{\hat B_{B_d}}F_{B_d}$ down to $5\%$. This implies an uncertainty of $10\%$ in
$\Delta M_{B_d}$ and  $\Delta M_{B_s}$ within the SM. On the other hand, the mixing induced
CP-asymmetries $S_{\psi\phi}$ and $S_{\psi K_S}$ have  much smaller hadronic uncertainties.

The hadronic uncertainties in the ratio $R_{\Delta M_B}$ are roughly at the $3\%$ level;
the theoretical error on the $b$ semileptonic CP-asymmetry $A^b_{sl}$ is around the $20\%$
level; the theoretical uncertainties in the rate of the $\bar B\to X_s\gamma$ decay are
below $10\%$; given the $5\%$ uncertainty on $F_{B_q}$ decay functions, the branching
ratio for $B^+\to\tau^+\nu$ has a theoretical error around the $10\%$ 
and a large parametric error due to $\vub$.

We stress that the situation
with other $B_i$ parameters, describing the hadronic
matrix elements of $\Delta F=2$ operators absent in the SM, is much worse.
Here a significant progress is desired.\\

On the experimental side, $\vep_K$, $\Delta M_{B_d}$,  $\Delta M_{B_s}$
and the ratio $R_{\Delta M_B}$ are very precisely measured with errors below the $1\%$ level.
$S_{\psi K_S}$ is known with an uncertainty of $\pm 3\%$ and the rate for the branching
ratio of the $\bar B\to X_s\gamma$ decay is known within $10\%$. On
the contrary, larger experimental uncertainties affect the measurements
of $S_{\psi\phi}$ in D0 and LHCb, which differ from one another by an order of magnitude,
but are still in agreement within the $2\sigma$-level, due to the large errors of the
single determinations. $A^b_{sl}$ has only been measured by D0 its experimental error is around the $20\%$. Similarly, $BR(B^+\to\tau^+\nu)$ is plagued by the same uncertainty.

\subsubsection{The CKM Matrix}\label{sec:ckmmatrix}
To evaluate the observables we need to  specify
the values of the CKM elements. As already stated in sec.~\ref{sec:MGF}, the CKM
matrix in this model is not unitary and is defined by
\begin{equation}
\tilde V = c_{u_L}\,V\,c_{d_L}\,,
\end{equation}
where $V$ is by construction a unitary $3\times3$ matrix and $c_{(u,d)_L}$ are the
cosines encoding the mixing between SM and exotic fermions. From eqs.~(\ref{FormulaSin1})
and (\ref{FormulaSin2}), we deduce that $c_{(u,d)_L}\approx1$, except for $t$ and $t'$.
As a result, within an excellent accuracy
the CKM matrix reads
\begin{equation}
\tilde V \simeq \left(
        \begin{array}{ccc}
            V_{ud} 					& V_{us} 					& V_{ub} \\
            V_{cd} 					& V_{cs}					& V_{cb}  \\
            c_{u_{L3}}\,V_{td} 	& c_{u_{L3}}\,V_{ts} 	&  c_{u_{L3}}\,V_{tb}\\
        \end{array}
\right)\,.
\label{Vtilde}
\end{equation}
In this  approximation, the deviation from the unitarity of
the CKM matrix is
\begin{equation}
  \big(\tilde V^\dagger\,\tilde V\big)_{ij}=\delta_{ij}-s^2_{u_{L3}}\,V^*_{ti}\,V_{tj}\,,\qquad\qquad
  \big(\tilde V\,\tilde V^\dagger\big)_{ij}=\delta_{ij}-s^2_{u_{L3}}\,\delta_{it}\,\delta_{jt}\,,
\end{equation}
The deviations are present only when the top-quark entries are considered
and are proportional to $s^2_{u_{L3}}$. All other entries of the CKM matrix coincide with the
corresponding entries of the unitary matrix $V$ up to negligible corrections.

The important implication of the latter finding is  that the  angle $\gamma$ in the unitary triangle is unaffected
by such deviations. In the approximation of eq.~(\ref{Vtilde}),
\begin{equation}
\tilde \gamma\equiv \arg\left(-\dfrac{\tilde V_{ud}\,\tilde V_{ub}^*}{\tilde V_{cd}\,\tilde V_{cb}^*}\right)=
\arg\left(-\dfrac{V_{ud}\,V_{ub}^*}{V_{cd}\,V_{cb}^*}\right)\,,
\end{equation}
and thus $\gamma$ does not depend on $c_{u_{L3}}$ or $s_{u_{L3}}$.

We state now how we fix  the values of the CKM elements. From
the tree-level experimental determinations of $\vus$, $\vcb$, $\vub$ and $\gamma$,
we fix the corresponding parameters of $\tilde V$. In this way, also the corresponding
parameters of the $V$ matrix are univocally fixed and using the unitarity of $V$, we
evaluate all the other entries of $V$. With all entries of $V$ fixed we compute the 
masses and mixings of all fermions and flavour gauge bosons by means of
eqs.~(\ref{SeeSawMasses})--(\ref{FGBmasses}).
Finally, knowing $c_{u_{L3}}$, we also determine the elements of the third row of
$\tilde V$.

\subsubsection{The Parameter Space of the Model}
Having determined $V$ what remains
is the calculation of the spectrum and the couplings of NP particles. In principle
they are fixed once, in addition to the SM parameters, we fix the {\it seven} NP
couplings $\lambda_{u,d}^{(\prime)}\,,g_Q\,,g_U\,,g_D$ and the {\it two} mass parameters
$M_u$ and $M_d$ in eqs.~\eqref{eq:covariantderivatives} and  \eqref{eq:lagrangian}. Still,
their actual determination is subtle since the energy scale at which the see-saw
relations of eqs.~\eqref{SeeSawMasses} hold is a priori not known. We identify
this scale with the mass of the lightest flavour gauge boson.

We fix the spectrum and the see-saw scale iteratively using the condition
that all exotic masses are above $m_t(m_t)$. As a first step we evaluate the
see-saw relation at $m_t(m_t)$ to obtain a rough estimate of the masses of exotic fermions
and lightest gauge boson. With this initial spectrum we run the masses of the
SM fermions to the newly defined see-saw scale including all intermediate exotic
fermion thresholds. The evaluation of the see-saw relation corrects the NP
spectrum. We repeat the procedure until the values of exotic fermion masses and see-saw
scale no longer change. Lastly, we evolve the exotic fermion masses down to the EW
scale.

For the numerical analysis it is necessary to scan the parameter space of the
model. We choose $\lambda_{u,d}\in(\, 0,\,1.5\,]$ and all other couplings
$\{\lambda^\prime_{u,d},\,g_Q,\,g_U,\,g_D\}\in(\, 0,\,1.1\,]$ to stay in the
perturbative regime of the theory. The two mass parameters
are varied between $M_u\in[\,100\gev\,,\,1\tev\,]$ and $M_d\in[\,30\gev\,,250\gev\,]$
following the discussion in Ref.~\cite{Grinstein:2010ve}. Unphysical points of the
parameter space, namely cases with $s_{u,d}$ or $c_{u,d}$ larger than $1$, are not
considered. Larger $M_u$ and $M_d$ values decouple the NP from the SM and
are therefore phenomenologically irrelevant. With respect to the analysis of
Ref.~\cite{Grinstein:2010ve} we are scanning over all NP parameters, including
$\lambda^\prime_{u,d}$ and $g_Q,\,g_U,\,g_D$.

\subsection{Results}

To present the features of the MGF model we are discussing, we use $\vus$ and $\vcb$
at their central values in tab.~\ref{tab:input} and
\begin{equation}
\vub=3.38\times 10^{-3}\, \qquad\text{and}\qquad 
\gamma=68^\circ\,,
\end{equation}
which are among the favoured values within the SM when the experimental values 
of both $S_{\psi K_S}$ and $R_{\Delta M_B}$ are taken into account. 

With this CKM matrix we list in tab.~\ref{tab:predictionsexperiment} the central values
for the SM predictions of the observables under consideration together with their experimental
determinations.
\begin{table}[h!]
\renewcommand{\arraystretch}{1}\setlength{\arraycolsep}{1pt}
\center{
\begin{tabular}{|l||l|}
\hline
SM predictions for exclusive $|V_{ub}|$						&  Experimental values\\
\hline\hline
$\Delta M_{B_d} = 0.592 \,\text{ps}^{-1}$					&  $\Delta M_{B_d} = 0.507(4) \,\text{ps}^{-1}$\hfill\cite{Nakamura:2010zzi}\\
$\Delta M_{B_s} = 20.28 \,\text{ps}^{-1}$					&  $\Delta M_{B_s} = 17.77(12) \,\text{ps}^{-1}$\hfill\cite{Nakamura:2010zzi}\\	
$R_{\Delta M_B}= 2.92\times 10^{-2}$						&  $R_{\Delta M_B}=(2.85\pm0.03)\times 10^{-2}$\hfill\cite{Nakamura:2010zzi}\\
$S_{\psi K_S}= 0.671$												&  $S_{\psi K_S}= 0.673(23)$\hfill\cite{Nakamura:2010zzi}\\
$S_{\psi\phi}= 0.0354 $												&  $\phi_s^{\psi\phi}= 0.55^{+0.38}_{-0.36}$\hfill\cite{Giurgiu:2010is,Abazov:2011ry}\\
																				&  $\phi_s^{\psi\phi}= 0.03\pm0.16\pm0.07 $\hfill\cite{Koppenburg:PC}\\	
$\Delta M_K = 0.4627 \times 10^{-2}\,\text{ps}^{-1}$	&  $\Delta M_K= 0.5292(9)\times 10^{-2} \,\text{ps}^{-1}$\hfill\cite{Nakamura:2010zzi}	 \\ 
$|\eps_K|= 1.791\times 10^{-3}$								&  $|\eps_K|= 2.228(11)\times 10^{-3}$\hfill\cite{Nakamura:2010zzi}			 \\
$A^b_{sl}=-0.0233\times10^{-2}$								&  $A^b_{sl}=(-0.787\pm0.172\pm0.093)\times10^{-2}$\hfill\cite{Abazov:2011yk}\\
$BR(b\to s\gamma)=3.15 \times10^{-4}$					&  $BR(b\to s\gamma)=(3.55\pm0.24\pm0.09) \times10^{-4}$\hfill\cite{Nakamura:2010zzi}\\
$BR(B^+\to\tau^+\nu)=0.849\times10^{-4}$				&  $BR(B^+\to\tau^+\nu)=(1.65\pm0.34)\times10^{-4}$\hfill\cite{Nakamura:2010zzi}\\
$R_{BR/\Delta M}=1.43\times 10^{-4}\ps$					&  $R_{BR/\Delta M}=(3.25\pm0.67)\times 10^{-4}\ps$\\
\hline
\end{tabular}} 
\caption{
  The SM predictions for the observables we shall consider using the exclusive determination of $|V_{ub}|$
  and the corresponding experimental values.
\label{tab:predictionsexperiment}}
\end{table}

Comparing these results with the data we make the following observations 
\begin{itemize}
\item[-]
$|\varepsilon_K|$ is smaller than its experimental determination,
while $S_{\psi K_S}$ is very close to the central experimental value, as it should be
for the chosen $\vub$ and $\gamma$.  
\item[-]
The mass differences $\Delta M_{B_d}$ and
$\Delta M_{B_s}$ are visibly above the data; also their ratio $R_{\Delta M_B}$ is above the
experimental determination, but in agreement at the $3\sigma$ level. 
\item[-]
$BR(B^+\to\tau^+\nu)$ is well below the data and consequently also the 
ratio $R_{BR/\Delta M}$ turns out to be below the measured central value by more than a factor of two. 
Even if the experimental error in $BR(B^+\to\tau^+\nu)$ is large, the parameter space of the model is strongly constrained. Furthermore, from the correlation among $R_{\Delta M_B}$ and $R_{BR/\Delta M}$ it is evident that the model can only deteriorate  the SM tension in these observables.
\item [-]
Concerning $S_{\psi\phi}$, the predicted value is consistent with the most 
recent data from CDF, D0 and LHCb.
\item[-]
$A^b_{sl}$ is well below the D0  data.
\item[-]
Finally the predicted central value for $BR(\bar B\to X_s\gamma)$ is
smaller than the central experimental value but consistent with it 
within the $2\sigma$ error range.
\end{itemize}

Any NP model that aims to remove or soften the anomalies listed above should simultaneously:
\begin{enumerate}
\item
Enhance $|\varepsilon_K|$ by roughly $20\%$ without affecting significantly $S_{\psi K_S}$.
\item
Suppress $\Delta M_{B_d}$ and $\Delta M_{B_s}$ by roughly $15\%$ and $10\%$, 
respectively.
\item
Slightly suppress $R_{\Delta M_{B}}$ by $3\%$.
\item
Strongly enhance $R_{BR/\Delta M}$ by $130\%$.
\item
Moderately enhance the value $BR(\bar B\to X_s\gamma)$ by $5-10\%$.
\end{enumerate}

As we shall see below, the model naturally satisfies requirements 1., 3. and 5. On the other side, it fails in 2. and 4.: indeed the mass differences $\Delta M_{B_{d,s}}$ can only be enhanced with respect to the corresponding would-be SM value; this enhancement is predicted to be significant if ones requires to solve the $|\varepsilon_K|$-$S_{\psi K_S}$ anomaly. Furthermore, the predicted value for the $R_{BR/\Delta M}$ can only be decreased, resulting in a tension on this observables more serious than in the SM.

In what follows we will look closer at the pattern of flavour violations in 
the MGF still keeping the input parameters at their central values. Subsequently 
we will comment on how some of our statements are softened when hadronic uncertainties 
in the input parameters are taken into account.

Considering now the NP contributions within MGF, we find the following 
pattern of effects:
\begin{itemize}
\item[-] 
If we neglect the contributions of flavour gauge bosons and of exotic quarks,
i.e. considering only the would-be SM contributions, the mixing amplitudes $M^i_{12}$
and $BR(\bar B\to X_s\gamma)$ are reduced with respect to the SM ones due to the
modification of the CKM matrix, encoded in the mixings $c_{{u,d}_L}$. As a result,
once the third-row entries of the CKM matrix are involved, the would-be SM values
of the considered observables are smaller than the values reported in tab.~\ref{tab:predictionsexperiment}.
\item[-]
The RR flavour gauge boson contributions are negligible for all observables 
in all the parameter space. We shall not consider such contributions in the
following description.
\item[-]
$|\varepsilon_K|$ is uniquely enhanced by the new box-diagram contributions involving exotic quarks,
while it is uniquely suppressed by heavy gauge flavour boson contributions.
Among the latter, the $LR$ contributions are the dominant ones, while
the $LL$ ones are safely negligible.
\item[-]
$\Delta M_{B_{d,s}}$ are also uniquely enhanced by the new
box-diagram contributions, but are mostly unaffected by heavy
flavour gauge boson contributions. This is in particular true for $\Delta M_{B_d}$,
while for $\Delta M_{B_s}$ the latter contributions can be non-negligible 
either enhancing or suppressing it. This is best appreciated
when considering the ratio $R_{\Delta M_B}$: this observable does
not show any dependence on the new box-diagram contributions, since in MGF the operator structure in box-diagram contributions
does not change with respect to the SM and the NP effects are the same in the $B_d$ and $B_s$ systems. As a result
any NP effect in this ratio should be attributed to the heavy gauge
flavour boson contributions, both $LL$ and $LR$.
\item[-]
The mixing induced CP-asymmetries $S_{\psi K_S}$ and
$S_{\psi\phi}$ are unaffected by the new box-diagram contributions. Similarly to $R_{\Delta M_B}$, this allows to see transparently
the heavy gauge flavour boson contributions, which was much 
harder in the case of $\Delta M_{B_{d,s}}$. We find that 
$S_{\psi K_S}$ is only affected by $LL$ contributions and can only be suppressed. $S_{\psi\phi}$
depends on both $LL$ and $LR$ contributions. Interestingly, the NP 
contributions interfere destructively with the SM contribution such that 
the sign of  $S_{\psi\phi}$ can in principle be reversed in this model. 
Similar conclusions hold for $A^b_{sl}$: it is not affected
by box-diagram contributions, the $LR$ contributions are almost completely negligible and
the $LL$ ones are the only relevant enhancing $|A^b_{sl}|$ towards the central value
of the experimental determination.
\item[-]
Finally, the branching ratio of $\bar B\to X_s\gamma$  can be significantly  affected
by the modifications in the SM magnetic penguin contributions 
that can only enhance this observable,
as already pointed out in Ref.~\cite{Grinstein:2010ve}. The heavy
gauge flavour boson contributions are negligible as discussed in
Ref.~\cite{Buras:2011zb}.
\end{itemize}
Having listed the basic characteristic of NP contributions in this 
model we will now present our numerical results in more detail stressing 
the important role of correlations among various observables identified 
in this model by us for the first time.

\subsubsection{Correlations Among the Observables}

In this section we discuss correlations among the observables. They will
allow us to constrain the parameter space of the model and see
whether this model is able or not to soften, or even solve,
the anomalies in the flavour data.

\begin{figure}[h!]
  \begin{center}
    \subfloat[Exclusive $V_{ub}$]{\label{fig:eK_SpsiKs_exclusive}\includegraphics[height=5cm]{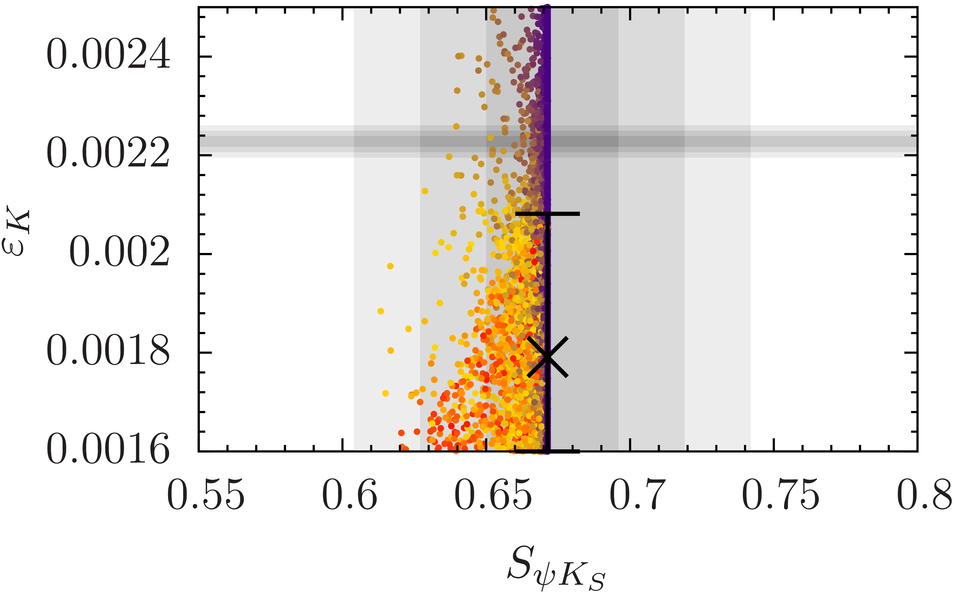}}
    ~
    \subfloat[Inclusive $V_{ub}$]{\label{fig:eK_SpsiKs_inclusive}\includegraphics[height=5cm]{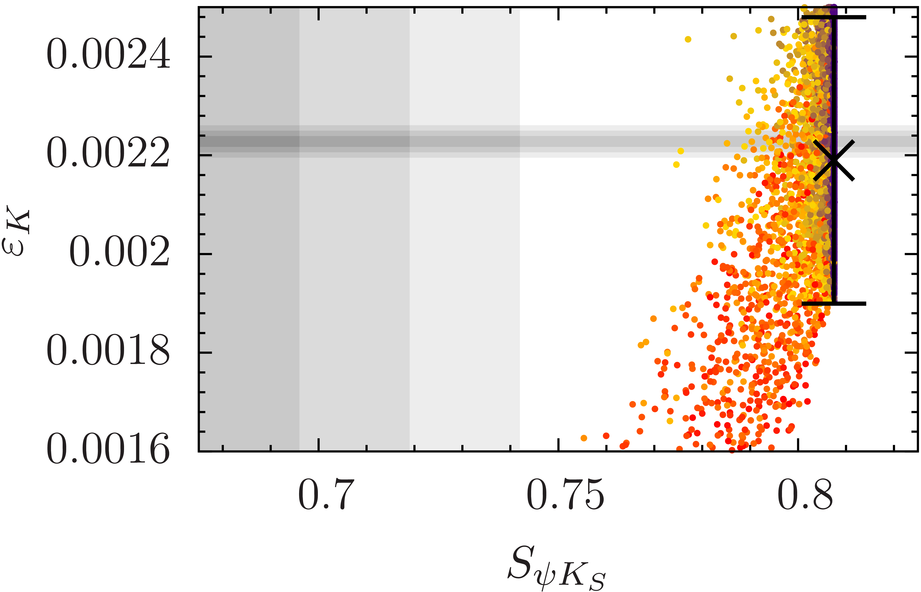}}
  \end{center}
  \caption{\it  
  The correlation of $\vep_K$ and $S_{\psi K_S}$. The shaded grey regions are the experimental $1\sigma$-$3\sigma$
  error ranges, while the cross is the central SM values reported
  in tab.~\ref{tab:predictionsexperiment}. The colour of the points represent the percentage of the
  box-diagram contributions (purple) and of the flavour gauge boson ones (red) in $\vep_K$. In the 
  NP points the theoretical error on $\vep_K$ is included. 
  \label{fig:eK_SpsiKs}}
\end{figure}
In fig.~\ref{fig:eK_SpsiKs_exclusive}, we show the correlation between
$\vep_K$ and $S_{\psi K_S}$. The plot confirms that the exclusive value of $|V_{ub}|$ is
favoured in this model. Indeed the NP contributions are able to solve the
$\vep_K$--$S_{\psi K_S}$ anomaly in a reasonably large region of the parameters space. This
happens when the $\vep_K$ prediction approaches the data due to  box-diagram
contributions (purple points), while $S_{\psi K_S}$ is mostly unaffected. This is
in particular possible  when the flavour gauge bosons contributions are negligible. 
 When the flavour gauge boson contributions are significant (red points) $S_{\psi K_S}$ 
is uniquely suppressed relatively to the SM value. However, as seen in the figure 
a combination of large box contributions as well as flavour gauge boson contributions
(yellow points) allows bringing $\vep_K$ in agreement with the data while keeping 
$S_{\psi K_S}$ within the $2\sigma$ experimental error range.

On the other hand, points for which the box contributions are negligible
and instead the flavour gauge boson contributions dominate in $\varepsilon_K$ (purely red points)
cannot explain the observed value of $\vep_K$. However, this kind of contributions are best suited
for the case with the inclusive determination of $|V_{ub}|$, reported in fig.~\ref{fig:eK_SpsiKs_inclusive}.
Still, there exist no points, which simultaneously bring $\vep_K$ and $S_{\psi K_S}$ in a $3\sigma$
agreement. The flavour gauge bosons contributions are not large enough to
suitably correct $S_{\psi K_S}$. We conclude that the inclusive determination of $|V_{ub}|$
is disfavoured in this model and we shall not further pursue this case.

\begin{figure}[h!]
  \begin{center}
  \subfloat[$F_{B_d}=205\mev$ ]{\label{fig:eK_DeltaMB_a}\includegraphics[height=5cm]{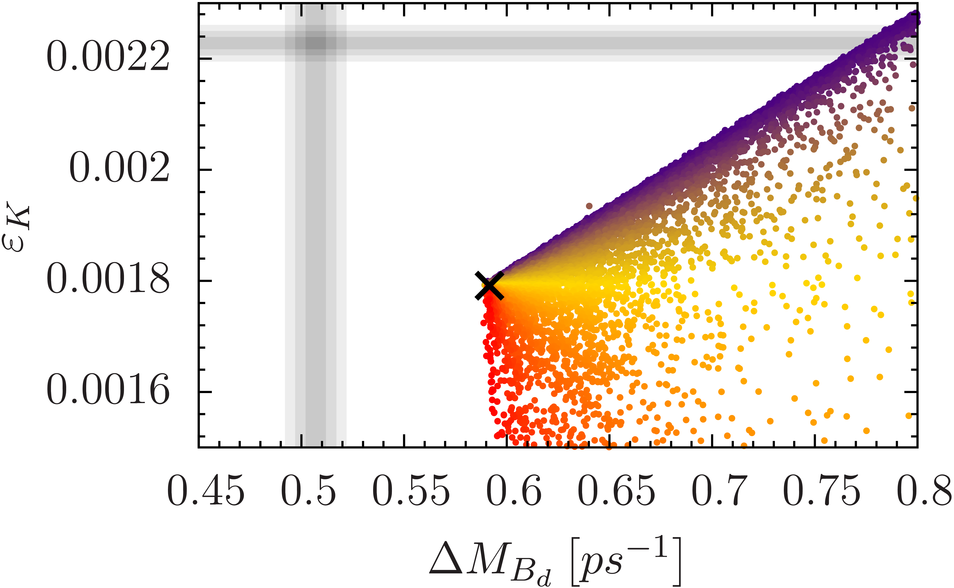}}~
  \subfloat[$F_{B_s}=250\mev$ ]{\label{fig:eK_DeltaMB_b}\includegraphics[height=5cm]{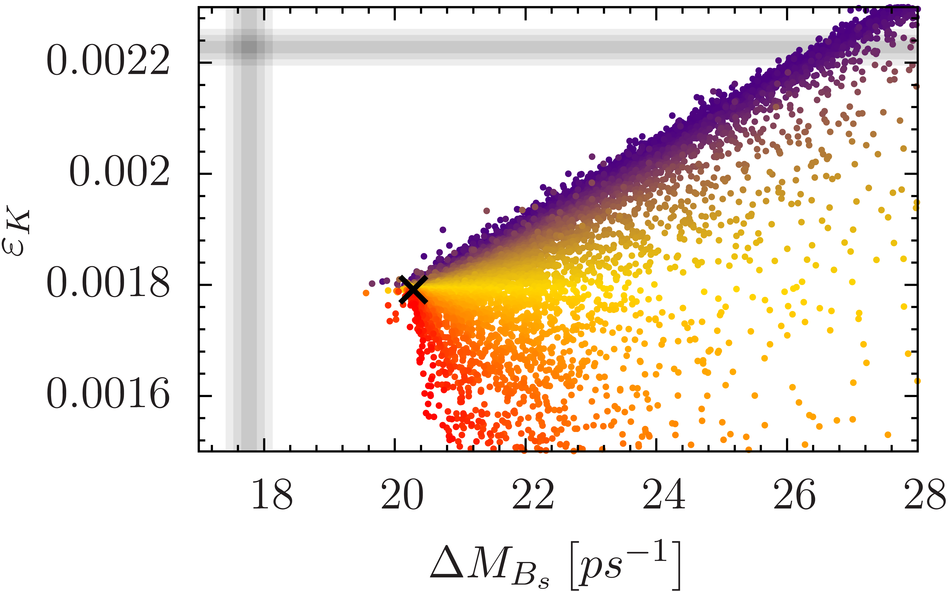}}\\[1em]
  \subfloat[$F_{B_d}=175\mev$ ]{\label{fig:eK_DeltaMB_c}\includegraphics[height=5cm]{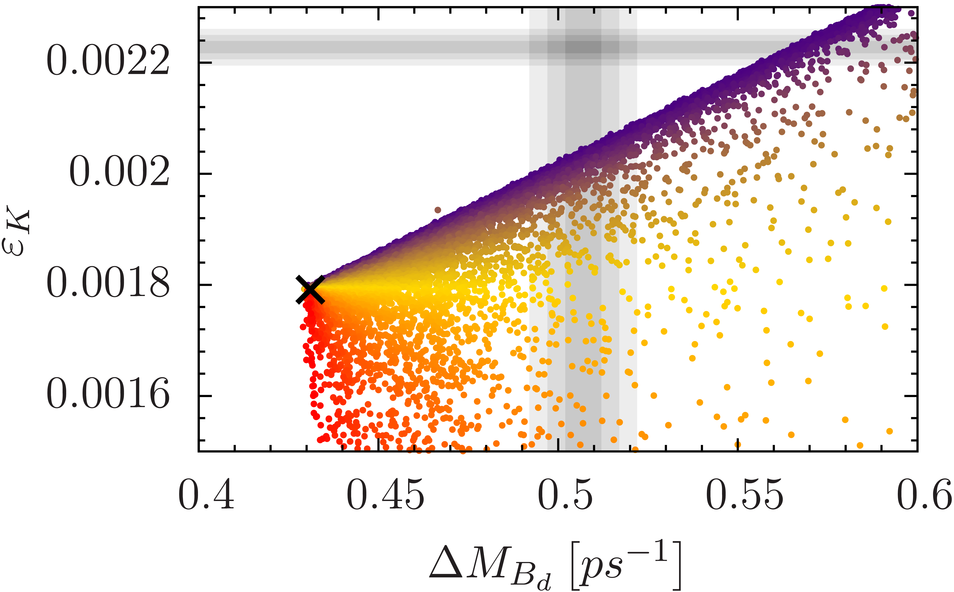}}~
  \subfloat[$F_{B_s}=225\mev$ ]{\label{fig:eK_DeltaMB_d}\includegraphics[height=5cm]{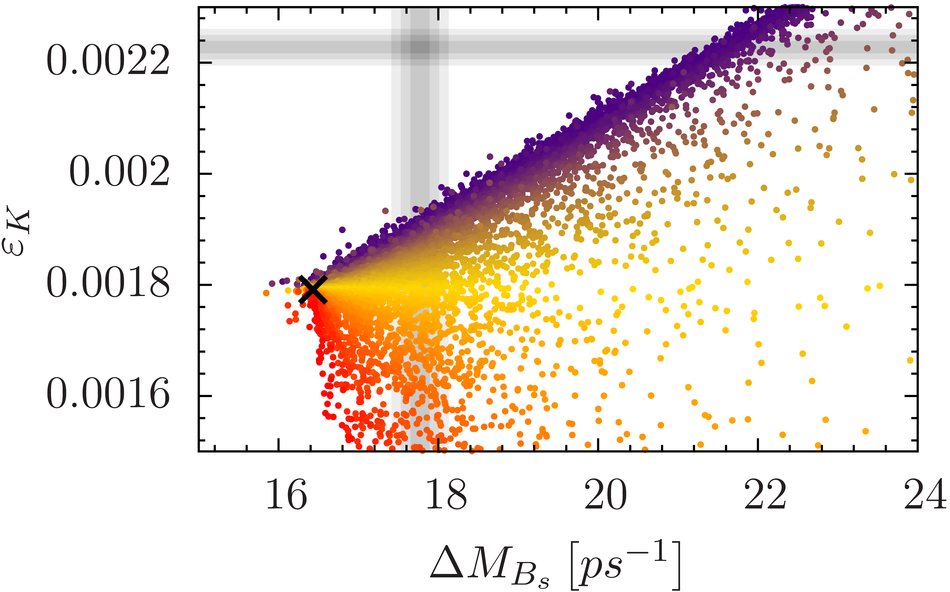}}
  \end{center}
  \caption{\it Correlation plot of $\vep_K$ with $\Delta M_{B_d}$ and $\Delta M_{B_s}$
  on the left and right, respectively. In the upper plots, we use for $F_{B_{d,s}}$ 
  the values reported in tab.~\ref{tab:input}, while for the lower plots we adopt
  smaller values: $F_{B_d}$ is reduced down to $85\%$ of its value, close to the $3\sigma$
  error level and $F_{B_s}$ is taken to be the new determination reported in
  Ref.~\cite{McNeile:2011ng}. In the plots above we have not included the $1\sigma$ error in $\varepsilon_K$
  to best illustrate the interplay of box- (purple) and tree-contributions (red).}
  \label{fig:eK_DeltaMB}
\end{figure}

In fig.~\ref{fig:eK_DeltaMB}, we present the correlations between $\vep_K$ and $\Delta M_{B_{d,s}}$. 
From \subref{fig:eK_DeltaMB_a} and \subref{fig:eK_DeltaMB_b}, we conclude that the model
cannot solve the $|\vep_K|-S_{\psi K_S}$ anomaly present in the SM, without worsening the already 
moderate agreement of this model with the $\Delta M_{B_{d,s}}$ experimental data. Indeed the values $\Delta M_{B_d}\approx0.75/ps$ and $\Delta M_{B_s}\approx27/ps$ are so larger that in the case that the central values of the weak decay 
constants $F_{B_{d,s}}$ do not change in the future, but their and other input parameter uncertainties are further reduced, we will 
have to conclude that the model fails to describe the $\Delta F=2$ data.

However, we
should emphasise that this problem could be avoided if the values for the weak decay constants
$F_{B_{d,s}}$ are smaller than the ones used in the plots \subref{fig:eK_DeltaMB_a}
and \subref{fig:eK_DeltaMB_b}. Indeed, in \subref{fig:eK_DeltaMB_c} we adopt a $15\%$ reduced
value for $F_{B_d}$, close to its $3\sigma$ value, while in \subref{fig:eK_DeltaMB_d} the last
determination of $F_{B_s}$ reported in Ref.~\cite{McNeile:2011ng}. This input,
modifies the SM values to be
\begin{equation}
\Delta M_{B_d}=0.43 \ps^{-1}\, \qquad\text{and}\qquad  
\Delta M_{B_s}=16.4 \ps^{-1}\,,
\end{equation}
such that now the enhancements of these observables by NP 
is welcomed by the data.
From plots \subref{fig:eK_DeltaMB_c} and \subref{fig:eK_DeltaMB_d} we deduce that
the NP contributions and the requirement of agreement of $|\varepsilon_K|$ with data within MGF
automatically enhance $\Delta M_{B_{d,s}}$. Even if also in this case  $\Delta M_{B_{d,s}}$
are found above the data, the model is performing much better than in the previous case. This 
exercise shows that on one hand it is crucial to get a better control over hadronic
parameters in order to obtain a clearer picture of NP contributions and on the other hand
that other more precise observables should be analysed until the uncertainties on
$\Delta M_{B_{d,s}}$ are lowered.

\begin{figure}[h!]
  \begin{center}
    \includegraphics[width=8cm]{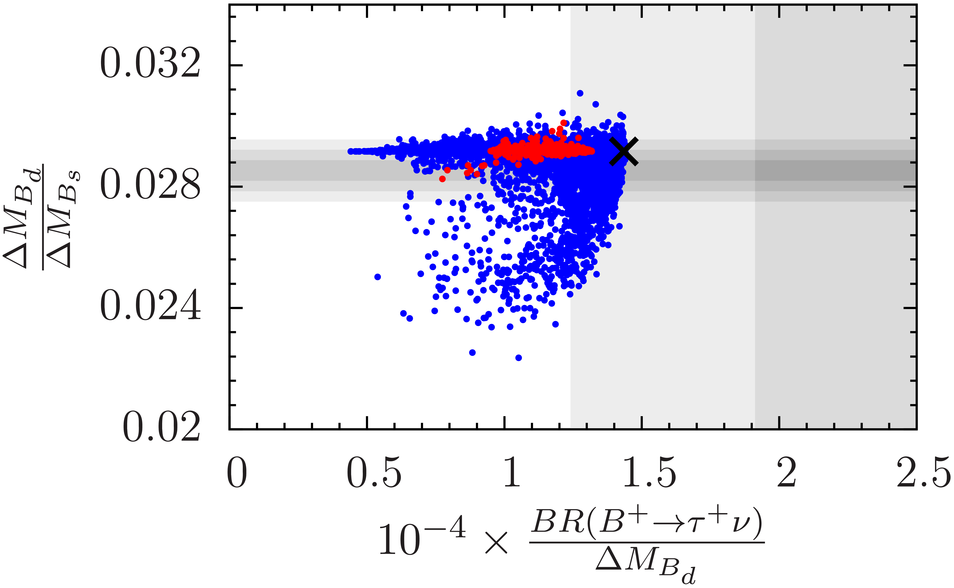}~
    \includegraphics[width=8cm]{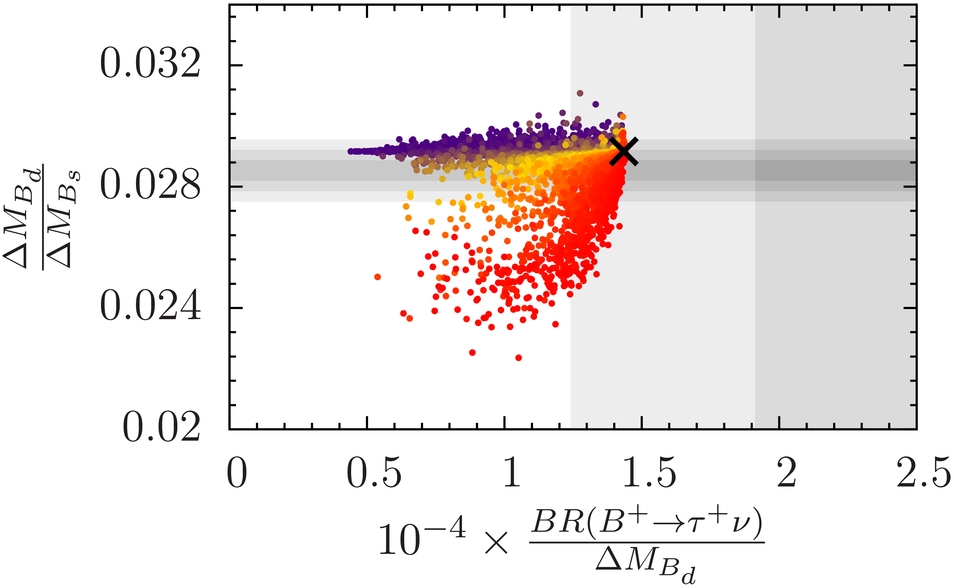}
  \end{center}
  \caption{\it Correlation plot for $R_{\Delta M_B}$ and $R_{BR/\Delta M}$. The grey regions
  refer to the experimental $1\sigma$-$3\sigma$  and $2\sigma$-$3\sigma$ error ranges for 
  $R_{\Delta M_B}$ and $R_{BR/\Delta M}$, respectively.
  The big black point refers to the SM values reported in tab.~\ref{tab:predictionsexperiment}. 
  On the left, red (blue) points refer to agreement (disagreement) of the points
  prediction of $\vep_K$ and the data at $3\sigma$ level.
  On the right the colours represent contribution of boxes and trees in $\varepsilon_K$.} 
  \label{fig:RDelta_RBRDelta}
\end{figure}

In fig.~\ref{fig:RDelta_RBRDelta}, we show the correlation between the ratio of the mass differences,
$R_{\Delta M_B}$, and the ratio among the branching ratio of the $B^+\to\tau^+\nu$ decay and the $\Delta M_{B_d}$,
$R_{BR/\Delta M}$. Both observables have negligible theoretical uncertainties
and are therefore very useful to provide strong constraints on the parameter space. 
On the left, the red points correspond to an $\vep_K$ prediction in agreement with the data at $3\sigma$,
while the blue ones  do not satisfy the $\vep_K$ constraint.
This plot largely constrains the parameter space of the model; 
for only very few points $\vep_K$, $R_{\Delta M_B}$ and $R_{BR/\Delta M}$ agree at $3\sigma$-level
with the data simultaneously.

Furthermore, all red points correspond to values for $R_{BR/\Delta M}$ smaller than the SM prediction and therefore the model can only worsen the SM tension. In the case that the experimental sensitivity to the $BR(B^+\to \tau^+\nu)$ improves, it will be possible to further constrain and possible exclude the present MGF model.

\begin{figure}[h!]
  \begin{center}
  \includegraphics[width=11cm]{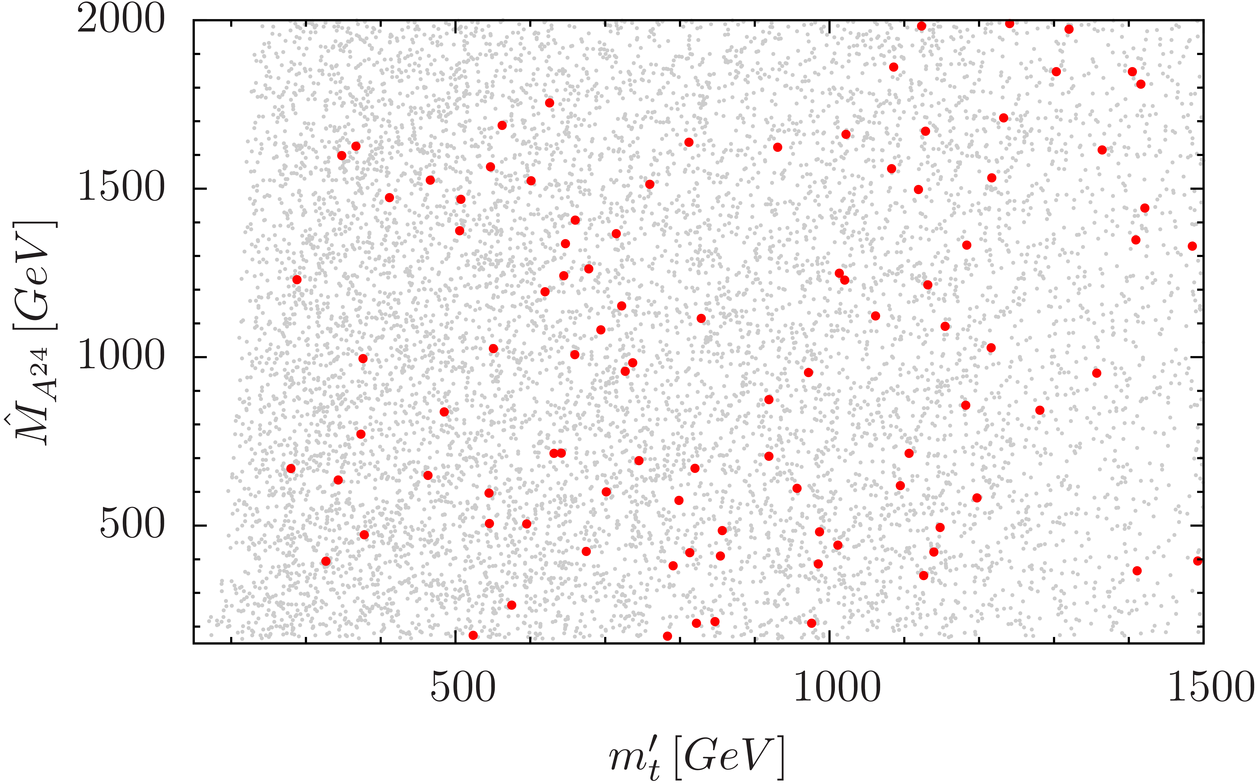}
  \end{center}
  \caption{\it $m_{t'}-\hat M_{A^{24}}$ parameter space. For all red points in the plot
  $R_{\Delta M_B}$, $R_{BR/\Delta M}$ and $\vep_K$ agree with the data at $3\sigma$-level.}
  \label{fig:mtopprime_MAH}
\end{figure}

Having reduced the parameter space, we concentrate now on a few other predictions of
the model. In fig.~\ref{fig:mtopprime_MAH}, we show the  $m_{t'}-\hat M_{A^{24}}$ parameter space,
where $m_{t'}$ is the mass of the exotic partner of the top-quark and $\hat M_{A^{24}}$ the
mass of the lightest neutral gauge boson; the corresponding particles  have the best chances to be 
detected at the LHC.
The red and blue points are those identified in fig.~\ref{fig:RDelta_RBRDelta}
to agree and disagree in  $R_{\Delta M_B}$, $R_{BR/\Delta M}$ and $\vep_K$
with the data at the $1\sigma$-$3\sigma$ level, respectively. 

Interestingly, the phenomenological results presented above hold not only for light but also for heavy $t^\prime$'s. Also the mass of the lightest flavour gauge boson is not bounded.

\begin{figure}[h!]
  \begin{center}
    \subfloat[]{\label{fig:SpsiphiAbsl}\includegraphics[height=5.4cm]{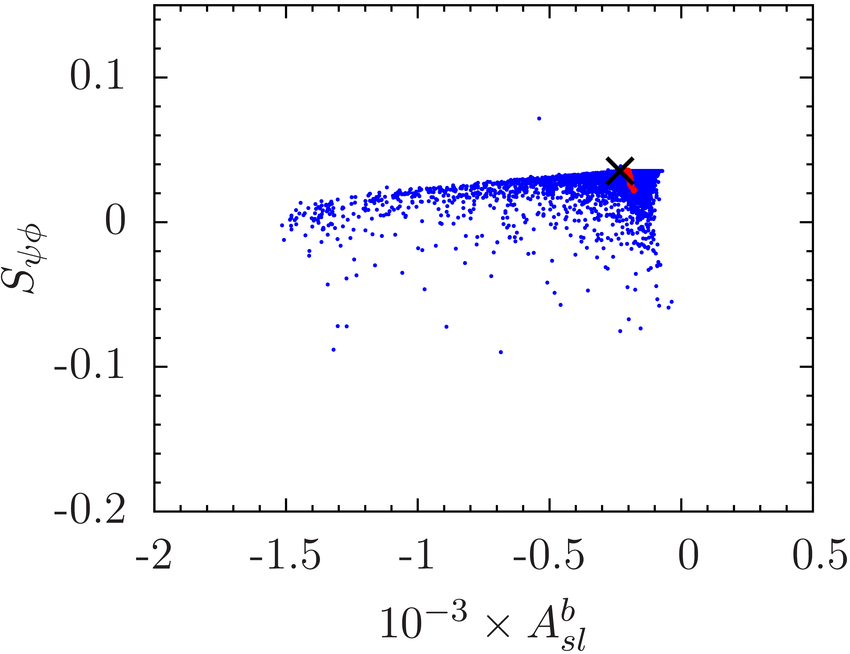}}
    \quad
    \subfloat[]{\label{fig:BSgammamtp}\includegraphics[height=5.5cm]{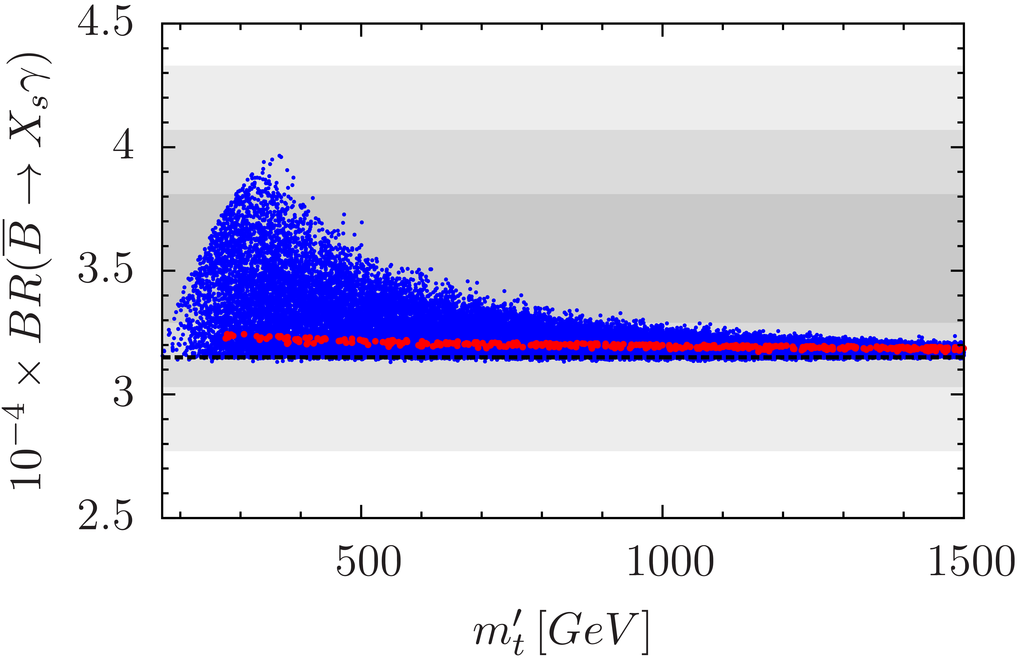}}
  \end{center}
  \caption{\it Correlation plot of $S_{\psi\phi}$ and $A^b_{sl}$ on the left and 
  $BR(\bar B\to X_s\gamma)$ and $m_t^\prime$ on the right. Grey regions refer to the
  experimental error ranges. The big black point refers to the SM values reported
  in tab.~\ref{tab:predictionsexperiment}. In red the points for which $R_{\Delta M_B}$, $R_{BR/\Delta M}$
  and $\vep_K$ agree with the data at $3\sigma$ level, in blue all others for which there is no agreement. 
  \label{fig:Predictions}}
\end{figure}

Furthermore, in fig.~\ref{fig:Predictions} we show two correlation plots which represent also
clear predictions for this model. Plot \subref{fig:SpsiphiAbsl} is the correlation for $S_{\psi\phi}$ and
$A^b_{sl}$ showing that only tiny deviations from the SM values are allowed: this turns out
to be an interesting result for $S_{\psi\phi}$, which is indeed close to the recent determination
of LHCb. On the other hand, $A^b_{sl}$ has only been measured by D0, but hopefully
LHCb will also have something to say in the near future. Once the experimental uncertainties are
lowered, such clear predictions will be essential to provide the final answer on how well
this model performs.

In plot \subref{fig:BSgammamtp}, we show the correlation of $BR(\bar B\to X_s\gamma)$ and $m_t^\prime$.
We confirm the finding of Grinstein {\it et al.} that in this model the NP contributions to
$BR(B\to X_s\gamma)$ always enhance it towards the central experiment value. However, interestingly 
only very small enhancements of this branching ratio are allowed when also the bounds from 
$\Delta F=2$ observables are taken into account.

%
%
\section{Comparison with other Models}
A complete comparison of the patterns of flavour violation in MGF with 
corresponding patterns found in numerous models \cite{Buras:2010wr} would 
require the study of $\Delta F=1$ processes, however already $\Delta F=2$ 
observables allow a clear distinction between the MGF and the simplest 
extensions of the SM. Here we just quote a few examples:
\begin{itemize}
\item[-]
In the original MFV framework restricted to LL operators, the 
so-called constrained MFV \cite{Blanke:2006ig}, the  $|\varepsilon_K|-S_{\psi K_S}$
anomaly can only be solved by enhancing $|\varepsilon_K|$ 
since in this framework $S_{\psi K_S}$ remains SM-like. In this framework 
then only the exclusive value of $\vub$ is viable. An example of such 
a framework is the model with a single universal extra dimension (UED) for 
which a very detailed analysis of $\Delta F=2$ observables has been 
performed in \cite{Buras:2002ej}. In fact this is a general property of CMFV models 
as demonstrated in \cite{Blanke:2006yh}. Thus after  $|\varepsilon_K|$ has been taken 
into account and contributions from tree-level heavy gauge boson 
exchanges have been eliminated MGF resembles CMFV if only $\Delta F=2$ 
processes are considered. However $\Delta F=1$ processes can provide 
a distinction. In fact whereas in MGF the NP contributions uniquely 
enhance $BR(\overline{B}\to X_s\gamma)$, in UED they uniquely suppress this branching 
ratio \cite{Buras:2003mk}. Concerning the  $|\varepsilon_K|$--$\Delta M_{B_{d,s}}$ tension MGF 
and CMFV are again similar.
\item[-]
The 2HDM framework with MFV and flavour blind phases, 
the so-called ${\rm 2HDM_{\overline{MFV}}}$ \cite{Buras:2010mh}, can on 
the other hand be easily distinguished from MGF. In this model NP 
contributions to $\varepsilon_K$ are tiny and the inclusive value 
of $\vub$ is required in order to obtain the correct value of $|\varepsilon_K|$. The interplay of the CKM phase with the flavour blind phases in Yukawa 
couplings and Higgs potential suppress $S_{\psi K_S}$ simultaneously 
enhancing the asymmetry $S_{\psi \phi}$. As in the case of MGF this asymmetry 
is SM-like or has a reversed sign. It is  $S_{\psi \phi}$ together with the 
value of $\vub$ which will distinguish MGF from ${\rm 2HDM_{\overline{MFV}}}$.
\item[-]
Finally, let us mention the left-right asymmetric model (LRAM) for which a very 
detailed 
FCNC analysis has been recently presented in \cite{Blanke:2011ry}. As this model has many free
parameters both values of $\vub$, inclusive and exclusive, are valid. 
The model contains many new phases and the 
$|\varepsilon_K|-S_{\psi K_S}$ anomaly can be solved in many ways. Moreover, 
the model struggles with the $\varepsilon_K$ constraint due to
huge neutral Higgs tree-level contributions. However, as demonstrated 
in Section  7 of that paper a simple structure of the right-handed mixing 
matrix gives a transparent solution to the 
$|\varepsilon_K|-S_{\psi K_S}$ anomaly by enhancing  $|\varepsilon_K|$, 
keeping $S_{\psi K_S}$ at the SM value and in contrast to MGF automatically 
{\it suppressing} $\Delta M_{B_{d,s}}$ and significantly {\it enhancing} $S_{\psi\phi}$. 
While MGF falls back in this comparison, one should emphasise than on the 
MGF has very few parameters and provides the explanation of quark 
masses and mixings, while this is not the case in the LRAM.
\end{itemize}

%
%
\section{Conclusion}
We have presented an extensive analysis of $\Delta F=2$ observables and $B\to X_s\gamma$
for a specific MGF model presented in Ref.~\cite{Grinstein:2010ve}, which is of special interest due to the small number of new parameters. In particular
we performed a detailed study of the effects of tree-level contributions due to the presence of heavy flavour gauge bosons.

Our main findings are as follows.
The model predicts a clear pattern of deviations from the SM:

\begin{itemize}
\item[-]
  Enhancements of  $|\varepsilon_K|$ and $\Delta M_{B_{d,s}}$ in a correlated manner 
by new box-diagram contributions and suppression of  $|\varepsilon_K|$ 
by tree-level heavy gauge boson contributions with only small impact on 
$\Delta M_{B_{d,s}}$.
\item[-]
Mixing induced CP-asymmetries  $S_{\psi K_S}$ and $S_{\psi \phi}$ are 
unaffected by box-diagram contributions, but receive sizeable destructive 
contributions from tree-level heavy gauge boson exchanges such that the sign of
$S_{\psi \phi}$ can be reversed. However, these effects are basically 
eliminated once the $\varepsilon_K$ constraint is taken into account.
\item[-]
The $\varepsilon_K-S_{\psi K_S}$ anomaly present in the SM is removed through 
the enhancement of $|\varepsilon_K|$, leaving $S_{\psi K_S}$ 
practically unmodified. This is achieved with the help of box-diagram contributions in the regions of the parameter space for which  they are dominant over heavy flavour gauge boson contributions, which interfere destructively with the SM amplitudes.
\item[-]
This structure automatically implies that in this model 
the exclusive determination of $\vub$ is favoured.
\item[-]
$b$ semileptonic CP-asymmetry
$A^b_{sl}$, that a
priori could receive large contributions from the tree-level flavour
gauge boson diagrams, remains close to the SM values once requiring $\vert\varepsilon_K\vert$
to be in agreement with the data. 
\item[-]
Most importantly, the  $\varepsilon_K$ constraint implies the 
central values of $\Delta M_{B_{d,s}}$ to be roughly $50\%$ higher than the very precise data. This disagreement 
cannot be cured fully by hadronic uncertainties although significant 
reduction in the values of $F_{B_{d,s}}$ could soften this problem.
\item[-]
We have pointed out that
the ratio of the mass differences, $\Delta M_{B_d}/\Delta M_{B_s}$ and the ratio
of the $B^+\to\tau^+\nu$ branching ratio and $\Delta M_{B_d}$, together
with $\varepsilon_K$, provide strong constraints on the parameter space of the model. Furthermore, the correlation among these two observables encodes a serious tension on the flavour data that can only be deteriorated in the model.
\item[-]
In agreement with Ref.~\cite{Grinstein:2010ve}, we find that $BR(B\to X_s\gamma)$ is
naturally enhanced in this model, bringing the theory closer to the data, still only
small corrections are allowed by the $\Delta F=2$ bounds.
\item[-]
We have demonstrated how this model can be distinguished by means of 
the flavour data from other extensions of the SM.
\end{itemize} 

In summary, the great virtue of this model is its predictivity, such that within the coming years it will be evident whether it can be considered as a valid description of low-energy data. Possibly the most transparent 
viability tests of the model are the future values of  $F_{B_{d,s}}$ 
and $\vub$. For the model to accommodate the data, $F_{B_{d,s}}$ have to be reduced 
by $15\%$ and $\vub$ has to be close to $3.4\times 10^{-3}$. Violation 
of any of these requirements will put this extension of the SM in trouble. 
We emphasise that the triple correlation between $\Delta M_{B_{d,s}}$, 
$\varepsilon_K$ and $S_{\psi K_S}$ was instrumental in reaching 
 this conclusion. The same result is derived by the complementary study on the $R_{BR/\Delta M}-R_{\Delta M_B}$ correlation. Our work shows that the study of correlations among flavour observables and the accuracy of non-perturbative parameters like $\hat B_K$ and  $F_{B_{d,s}}$ are crucial for the indirect searches for physics beyond the Standard Model.


\section*{Acknowledgements}

We would like to thank Benjam\'in Grinstein for useful details on the Ref.~\cite{Grinstein:2010ve}, Ulrich Nierste and Paride Paradisi for very interesting discussions. This research was done in the context of the ERC Advanced Grant project ''FLAVOUR''(267104). The work of MVC has been partially supported by the Graduiertenkolleg GRK 1054 of DFG.

%
%
\appendix
\mathversion{bold}
\section{Feynman Rules for MGF}
\mathversion{normal}

\mathversion{bold}
\subsection{Couplings of SM Gauge and Goldstone bosons}
\label{app:coup2smgauge}
\mathversion{normal}
\begin{multicols}{2}

\subsection*{$\gamma$ coupling}
\mathversion{normal}
 The photon coupling remains unchanged; proportional to the quark charges $Q_u$ and $Q_d$.
\mathversion{bold}
\subsection*{$G$ coupling}
\mathversion{normal} 

The gluon coupling remains unchanged; proportional to the colour generators $\lambda_{SU(3)}^a$.

\mathversion{bold}
\subsection*{$W^{\pm}$ coupling}  
\mathversion{normal}
\includegraphics{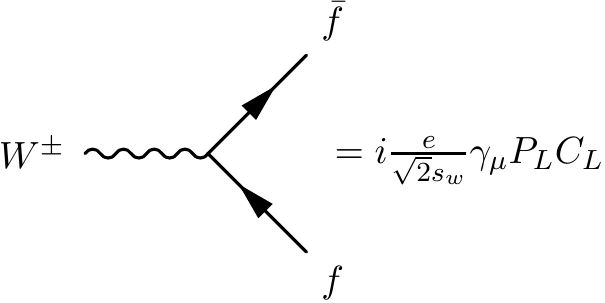}\\
 with the actual values for $\bar f$, $f$ and $C_L$:\\[1em]

 \begin{tabular}[h]{lp{0.5em}l}
   $W^+ \bar u_{i} d_{j}$ 	&:	& $C_L=c_{uLi} V_{ij} c_{dLj}$ \\
   $W^+ \bar u'_{i} d'_{j}$ 	&:	& $C_L=s_{uLi} V_{ij} s_{dLj}$ \\
   $W^+ \bar u_{i} d'_{j}$ 	&:	& $C_L=c_{uLi} V_{ij} s_{dLj}$ \\
   $W^+ \bar u'_{i} d_{j}$ 	&:	& $C_L=s_{uLi} V_{ij} c_{dLj}$ \\[2em]

   $W^- \bar d_{j} u_{i}$ 	&:	& $C_L=c_{uLi} V_{ij}^* c_{dLj}$ \\
   $W^- \bar d'_{j} u'_{i}$ 	&:	& $C_L=s_{uLi} V_{ij}^* s_{dLj}$ \\
   $W^- \bar d_{j} u'_{i}$ 	&:	& $C_L=s_{uLi} V_{ij}^* c_{dLj}$ \\
   $W^- \bar d'_{j} u_{i}$ 	&:	& $C_L=c_{uLi} V_{ij}^* s_{dLj}$
 \end{tabular}

\mathversion{bold}
\subsection*{$Z$ coupling}  
\mathversion{normal}
\includegraphics{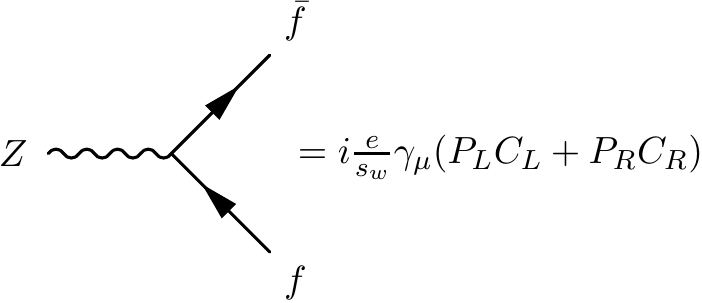}\\
 with the actual values for $\bar f$, $f$ and C:\\[1em]

 \begin{tabular}[h]{lp{0.5em}l}
   $Z \bar u_{i} u_{i}$ 	&:	& $C_R=- \frac{s_w^2}{c_w} Q_u$ \\
   				&:	& $C_L=\frac{T_{u}^3 c_{uLi}^2-s_w^2 Q_u}{c_w}$ \\
   $Z \bar u'_{i} u'_{i}$ 	&:	& $C_R=- \frac{s_w^2}{c_w} Q_u$ \\
   				&:	& $C_L=+\frac{T_{u}^3 s_{uLi}^2-s_w^2 Q_u}{c_w}$ \\
   $Z \bar u_{i} u'_{i}$ 	&:	& $C_R=0$ \\
   				&:	& $C_L=+\frac{T_{u}^3}{c_w} c_{uLi}s_{uLi}$ \\
   $Z \bar u'_{i} u_{i}$ 	&:	& $C_R=0$ \\
   				&:	& $C_L=+\frac{T_{u}^3}{c_w} c_{uLi}s_{uLi}$ \\[2em]

   $Z \bar d_{i} d_{i}$ 	&:	& $C_R=- \frac{s_w^2}{c_w} Q_d$ \\
   				&:	& $C_L=\frac{T_{d}^3 c_{dLi}^2-s_w^2 Q_d}{c_w}$ \\
   $Z \bar d'_{i} d'_{i}$ 	&:	& $C_R=- \frac{s_w^2}{c_w}  Q_d$ \\
   				&:	& $C_L=+\frac{T_{d}^3 s_{dLi}^2-s_w^2 Q_d}{c_w}$ \\
   $Z \bar d_{i} d'_{i}$ 	&:	& $C_R=0$ \\
   				&:	& $C_L=+\frac{T_{d}^3}{c_w} c_{dLi}s_{dLi}$ \\
   $Z \bar d'_{i} d_{i}$ 	&:	& $C_R=0$ \\
   				&:	& $C_L=+\frac{T_{d}^3}{c_w} c_{dLi}s_{dLi}$ \\
 \end{tabular}

\mathversion{bold}
\subsection*{$H$ coupling}  
\mathversion{normal}
\includegraphics{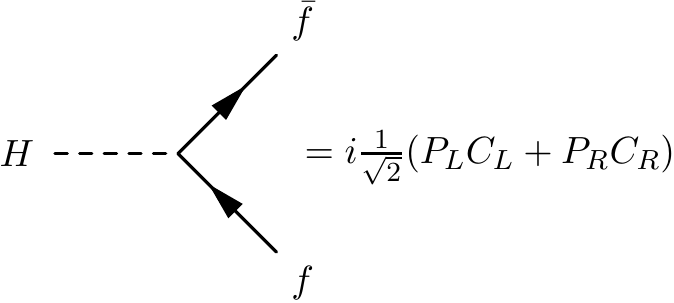}\\
with the actual values for $\bar f$, $f$ and C:\\[1em]

 \begin{tabular}[h]{lp{0.5em}l}
   $H \bar u_{i} u_{i}$ 	&:	& $C_R=C_L=+\lambda_u s_{uRi}c_{uLi}$\\
   $H \bar u'_{i} u'_{i}$ 	&:	& $C_R=C_L=-\lambda_u c_{uRi}s_{uLi}$\\
   $H \bar u_{i} u'_{i}$ 	&:	& $C_R=C_L=-\lambda_u c_{uRi}c_{uLi}$\\
   $H \bar u'_{i} u_{i}$ 	&:	& $C_R=C_L=+\lambda_u s_{uRi}s_{uLi}$\\[2em]

   $H \bar d_{i} d_{i}$ 	&:	& $C_R=C_L=+\lambda_d s_{dRi}c_{dLi}$\\
   $H \bar d'_{i} d'_{i}$ 	&:	& $C_R=C_L=-\lambda_d c_{dRi}s_{dLi}$\\
   $H \bar d_{i} d'_{i}$ 	&:	& $C_R=C_L=-\lambda_d c_{dRi}c_{dLi}$\\
   $H \bar d'_{i} d_{i}$ 	&:	& $C_R=C_L=+\lambda_d s_{dRi}s_{dLi}$\\
 \end{tabular}

\mathversion{bold}
\subsection*{$G_0$ coupling}  
\mathversion{normal}
\includegraphics{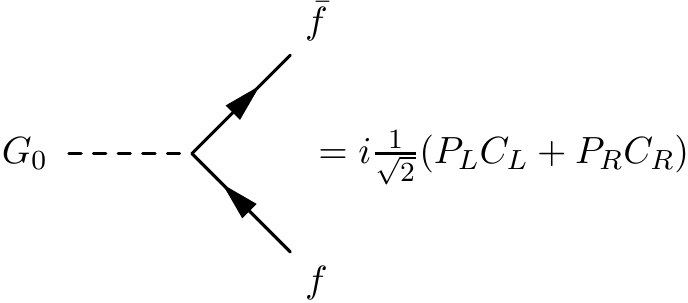}\\
 with the actual values for $\bar f$, $f$ and C:\\[1em]

 \begin{tabular}[h]{lp{0.5em}l}
   $G_0 \bar u_{i} u_{i}$ 	&:	& $C_R=-C_L=-i\lambda_u s_{uRi}c_{uLi}$\\
   $G_0 \bar u'_{i} u'_{i}$ 	&:	& $C_R=-C_L=+i\lambda_u c_{uRi}s_{uLi}$\\
   $G_0 \bar u_{i} u'_{i}$ 	&:	& $C_R=-C_L=+i\lambda_u c_{uRi}c_{uLi}$\\
   $G_0 \bar u'_{i} u_{i}$ 	&:	& $C_R=-C_L=-i\lambda_u s_{uRi}s_{uLi}$\\[2em]

   $G_0 \bar d_{i} d_{i}$ 	&:	& $C_R=-C_L=+i\lambda_d s_{dRi}c_{dLi}$\\
   $G_0 \bar d'_{i} d'_{i}$ 	&:	& $C_R=-C_L=-i\lambda_d c_{dRi}s_{dLi}$\\
   $G_0 \bar d_{i} d'_{i}$ 	&:	& $C_R=-C_L=-i\lambda_d c_{dRi}c_{dLi}$\\
   $G_0 \bar d'_{i} d_{i}$ 	&:	& $C_R=-C_L=+i\lambda_d s_{dRi}s_{dLi}$\\
 \end{tabular}

\mathversion{bold}
\subsection*{$G^{\pm}$ coupling}  
\mathversion{normal}
\includegraphics{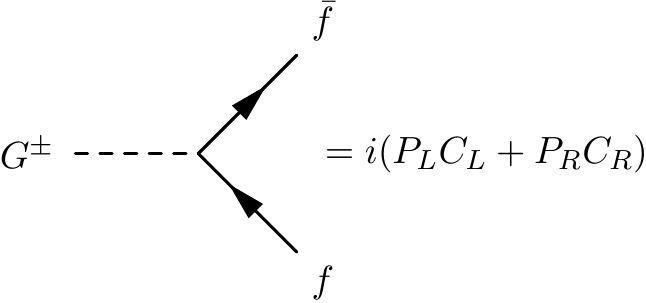}\\
 with the actual values for $\bar f$, $f$ and $C_L$:\\[1em]

 \begin{tabular}[h]{lp{0.5em}l}
   $G^+ \bar u_{i} d_{j}$ 	&:	& $C_R=+\lambda_d c_{uLi} V_{ij} s_{dRj}$   \\
   				&:	& $C_L=-\lambda_u s_{uRi} V_{ij} c_{dLj}$   \\
   $G^+ \bar u'_{i} d'_{j}$ 	&:	& $C_R=-\lambda_d s_{uLi} V_{ij} c_{dRj}$   \\
   				&:	& $C_L=+\lambda_u c_{uRi} V_{ij} s_{dLj}$   \\
   $G^+ \bar u_{i} d'_{j}$ 	&:	& $C_R=-\lambda_d c_{uLi} V_{ij} c_{dRj}$   \\
   				&:	& $C_L=+\lambda_u c_{uRi} V_{ij} c_{dLj}$   \\
   $G^+ \bar u'_{i} d_{j}$ 	&:	& $C_R=+\lambda_d s_{uLi} V_{ij} s_{dRj}$   \\
   				&:	& $C_L=-\lambda_u s_{uRi} V_{ij} s_{dLj}$   \\[2em]

   $G^- \bar d_{j} u_{i}$ 	&:	& $C_R=-\lambda_u s_{uRi} V_{ij}^* c_{dLj}$ \\
   				&:	& $C_L=+\lambda_d c_{uLi} V_{ij}^* s_{dRj}$ \\
   $G^- \bar d'_{j} u'_{i}$ 	&:	& $C_R=+\lambda_u c_{uRi} V_{ij}^* s_{dLj}$ \\
   				&:	& $C_L=-\lambda_d s_{uLi} V_{ij}^* c_{dRj}$ \\
   $G^- \bar d_{j} u'_{i}$ 	&:	& $C_R=+\lambda_u c_{uRi} V_{ij}^* c_{dLj}$ \\
   				&:	& $C_L=-\lambda_d c_{uLi} V_{ij}^* c_{dRj}$ \\
   $G^- \bar d'_{j} u_{i}$ 	&:	& $C_R=-\lambda_u s_{uRi} V_{ij}^* s_{dLj}$ \\
   				&:	& $C_L=+\lambda_d s_{uLi} V_{ij}^* s_{dRj}$ \\
 \end{tabular}

\end{multicols}
\mathversion{bold}
\subsection{Couplings of Flavour Gauge Bosons}
\label{app:coup2flavour}
\mathversion{normal}
  There are three types of flavour gauge bosons, $A_Q^a$, $A_U^a$, and $A_D^a$,
  which are flavour eigenstates but not mass eigenstates. We denote the mass
  eigenstates with $\hat{A}^m$, where $m=1,\dots,24$. 
  \begin{align*}
    \chi 	&= (A_Q^1,\dots,A_Q^8,A_U^1,\dots,A_U^{8},A_D^1,\dots,A_D^{8})^T\\
    \varphi  	&= (\hat{A}^1,\dots,\hat{A}^{24})^T.
  \end{align*}
  Flavour basis $\chi$ and mass basis $\varphi$ are connected through
  the transformation
  \begin{equation}
    \chi =\mathcal{W} \varphi\,,
    \label{}
  \end{equation}
  where $\mathcal{W}$ is obtained numerically by diagonalising the 
  mass matrix in eq.~\eqref{eq:massmatrix} such that:
  \begin{equation}
    \mathcal{\hat{M}}^2_A = \mathcal{W}^T \mathcal{M}^2_A \mathcal{W}
    \label{eq:diagmassmatrix}
  \end{equation}
  and $\mathcal{\hat{M}}_A$ is a diagonal mass-matrix.
  
  We define:
   \begin{equation*}
     U^T=( u,c,t,u',c',t')\quad \text{and}\quad
     D^T=( d,s,b,d',s',b')
   \end{equation*}
   such that the coupling of the flavour gauge bosons to the quarks are described
   by the Lagrangian-part
   \begin{align*}
     \mathcal{L}\supset	& + \bar U_i  \gamma_{\mu}(\cG^u_L + \cG^u_R)_{ij,m} U_j \cdot \chi_{m}\\
     			& + \bar D_i  \gamma_{\mu}(\cG^d_L + \cG^d_R)_{ij,m} D_j \cdot \chi_{m}
   \end{align*}
   where $m$ is understood to run from $1$ to $24$  and the tensors $\cG\equiv\cG[C_{L,R},\,g_{Q,U,D}]$
   can be read off from the couplings of flavour eigenstates $A_Q$, $A_U$, and
   $A_D$ to the quarks, that are listed below: i.e. 
   \begin{equation}
   \left(\cG_L^u\right)_{13,1}=\dfrac{g_Q}{2}\,c_{u_{L1}}\,(\lambda_{SU(3)}^1)_{13}\, c_{u_{L3}}\,,
   \qquad\qquad
   \left(\cG_R^d\right)_{42,18}=\dfrac{g_D}{2}\,s_{d_{L1}}\,(\lambda_{SU(3)}^2)_{12}\, c_{d_{L2}}\,.
   \label{DefGunhat}
   \end{equation}

   The rotation to the mass-eigenstates of the heavy gauge bosons redefines the
   couplings:
   \begin{align*}
     \mathcal{L}\supset	& + \bar U_i  \gamma_{\mu}(\hat \cG^u_L + \hat \cG^u_R)_{ij,k} U_j \cdot \varphi_{k}\\
     			& + \bar D_i  \gamma_{\mu}(\hat \cG^d_L + \hat \cG^d_R)_{ij,k} D_j \cdot \varphi_{k}\,,
   \end{align*}
   where
\begin{equation}
\left(\hat \cG^\alpha_{L,R}\right)_{ij,m}=\sum_{k} \cW(\varphi_m,\,\chi_k)\,\left(\cG^\alpha_{L,R}\right)_{ij,m}\,,
\qquad\text{with}\qquad
\alpha=u,d\,.
\label{DefGhat}
\end{equation}
Notice that in the rest of the paper we either use $\{i,\,j\}\in\{1,\ldots,6\}$ or refer directly to the (SM or exotic) quark flavour: i.e.
\begin{equation}
\left(\hat \cG^u_{L,R}\right)_{15,m}\equiv\left(\hat \cG^u_{L,R}\right)_{uc',m}\,,\qquad\qquad
\left(\hat \cG^d_{L,R}\right)_{23,m}\equiv\left(\hat \cG^d_{L,R}\right)_{sb,m}\,.
\end{equation}

\begin{multicols}{2}
\mathversion{bold}
  \subsection*{$A_Q$ coupling}
\mathversion{normal}
\includegraphics{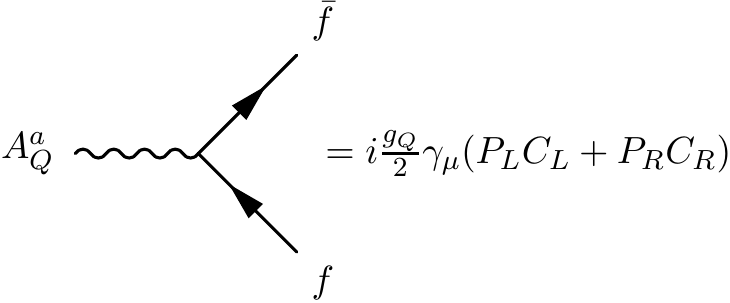}\\
   with the actual values for $\bar f$, $f$ and C:\\[1em]
   \\
   \begin{tabular}[h]{lp{0.5em}l}
     $A_Q^a \bar u_{i}  u_{j}$ 	&:	& $C_R=+s_{uRi}(V\lambda_{SU(3)}^aV^\dagger)_{ij}s_{uRj}$ \\
     				&:	& $C_L=+c_{uLi}(V\lambda_{SU(3)}^aV^\dagger)_{ij}c_{uLj}$ \\
     $A_Q^a \bar u'_{i} u'_{j}$	&:	& $C_R=+c_{uRi}(V\lambda_{SU(3)}^aV^\dagger)_{ij}c_{uRj}$ \\
     				&:	& $C_L=+s_{uLi}(V\lambda_{SU(3)}^aV^\dagger)_{ij}s_{uLj}$ \\
     $A_Q^a \bar u_{i}  u'_{j}$	&:	& $C_R=-s_{uRi}(V\lambda_{SU(3)}^aV^\dagger)_{ij}c_{uRj}$ \\
     				&:	& $C_L=+c_{uLi}(V\lambda_{SU(3)}^aV^\dagger)_{ij}s_{uLj}$ \\
     $A_Q^a \bar u'_{i} u_{j}$ 	&:	& $C_R=-c_{uRi}(V\lambda_{SU(3)}^aV^\dagger)_{ij}s_{uRj}$ \\
     				&:	& $C_L=+s_{uLi}(V\lambda_{SU(3)}^aV^\dagger)_{ij}c_{uLj}$ \\[2em]

     $A_Q^a \bar d_{i}  d_{j}$ 	&:	& $C_R=+s_{dRi}(\lambda_{SU(3)}^a)_{ij}s_{dRj}$ \\
     				&:	& $C_L=+c_{dLi}(\lambda_{SU(3)}^a)_{ij}c_{dLj}$ \\
     $A_Q^a \bar d'_{i} d'_{j}$	&:	& $C_R=+c_{dRi}(\lambda_{SU(3)}^a)_{ij}c_{dRj}$ \\
     				&:	& $C_L=+s_{dLi}(\lambda_{SU(3)}^a)_{ij}s_{dLj}$ \\
     $A_Q^a \bar d_{i}  d'_{j}$	&:	& $C_R=-s_{dRi}(\lambda_{SU(3)}^a)_{ij}c_{dRj}$ \\
     				&:	& $C_L=+c_{dLi}(\lambda_{SU(3)}^a)_{ij}s_{dLj}$ \\
     $A_Q^a \bar d'_{i} d_{j}$ 	&:	& $C_R=-c_{dRi}(\lambda_{SU(3)}^a)_{ij}s_{dRj}$ \\
     				&:	& $C_L=+s_{dLi}(\lambda_{SU(3)}^a)_{ij}c_{dLj}$ \\ 
   \end{tabular}

\mathversion{bold}
  \subsection*{$A_U$ coupling}
\mathversion{normal}
\includegraphics{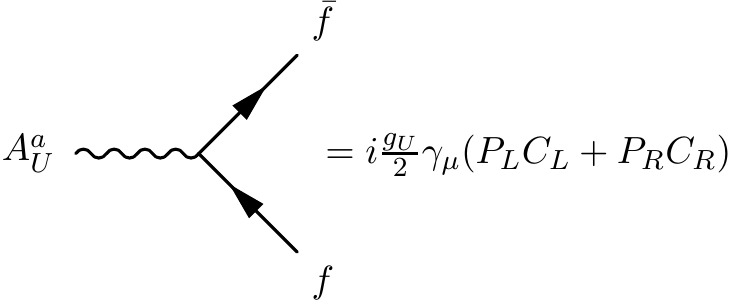}\\
   with the actual values for $\bar f$, $f$ and C:\\[1em]
   
   \begin{tabular}[h]{lp{0.5em}l}
     $A_U^a \bar u_{i}  u_{j}$ 	&:	& $C_R=+c_{uRi}(\lambda_{SU(3)}^a)_{ij}c_{uRj}$	\\
     				&:	& $C_L=+s_{uLi}(\lambda_{SU(3)}^a)_{ij}s_{uLj}$	\\
     $A_U^a \bar u'_{i} u'_{j}$	&:	& $C_R=+s_{uRi}(\lambda_{SU(3)}^a)_{ij}s_{uRj}$ \\
     				&:	& $C_L=+c_{uLi}(\lambda_{SU(3)}^a)_{ij}c_{uLj}$ \\
     $A_U^a \bar u_{i}  u'_{j}$	&:	& $C_R=+c_{uRi}(\lambda_{SU(3)}^a)_{ij}s_{uRj}$	\\
     				&:	& $C_L=-s_{uLi}(\lambda_{SU(3)}^a)_{ij}c_{uLj}$ \\
     $A_U^a \bar u'_{i} u_{j}$ 	&:	& $C_R=+s_{uRi}(\lambda_{SU(3)}^a)_{ij}c_{uRj}$	\\
     				&:	& $C_L=-c_{uLi}(\lambda_{SU(3)}^a)_{ij}s_{uLj}$ 
   \end{tabular}

\mathversion{bold}
  \subsection*{$A_D$ coupling}
\mathversion{normal}
\includegraphics{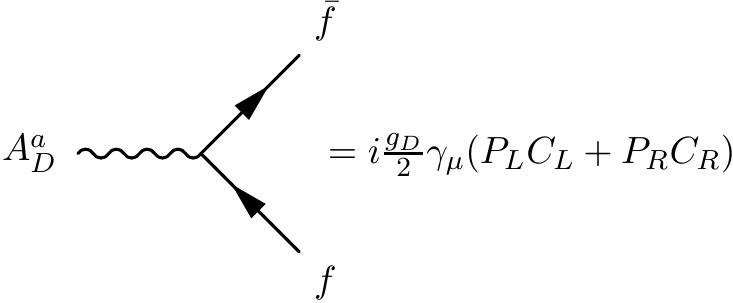}\\
   with the actual values for $\bar f$, $f$ and C:\\[1em]

   \begin{tabular}[h]{lp{0.5em}l}
     $A_D^a \bar d_{i}  d_{j}$ 	&:	& $C_R=+c_{dRi}(\lambda_{SU(3)}^a)_{ij}c_{dRj}$	\\
     				&:	& $C_L=+s_{dLi}(\lambda_{SU(3)}^a)_{ij}s_{dLj}$	\\
     $A_D^a \bar d'_{i} d'_{j}$	&:	& $C_R=+s_{dRi}(\lambda_{SU(3)}^a)_{ij}s_{dRj}$ \\
     				&:	& $C_L=+c_{dLi}(\lambda_{SU(3)}^a)_{ij}c_{dLj}$ \\
     $A_D^a \bar d_{i}  d'_{j}$	&:	& $C_R=+c_{dRi}(\lambda_{SU(3)}^a)_{ij}s_{dRj}$	\\
     				&:	& $C_L=-s_{dLi}(\lambda_{SU(3)}^a)_{ij}c_{dLj}$ \\
     $A_D^a \bar d'_{i} d_{j}$ 	&:	& $C_R=+s_{dRi}(\lambda_{SU(3)}^a)_{ij}c_{dRj}$	\\
     				&:	& $C_L=-c_{dLi}(\lambda_{SU(3)}^a)_{ij}s_{dLj}$ 
   \end{tabular}

\end{multicols}

\section{Couplings of the Lightest Flavour Gauge Boson}
\begin{figure}
  \begin{center}
    \includegraphics[width=7cm]{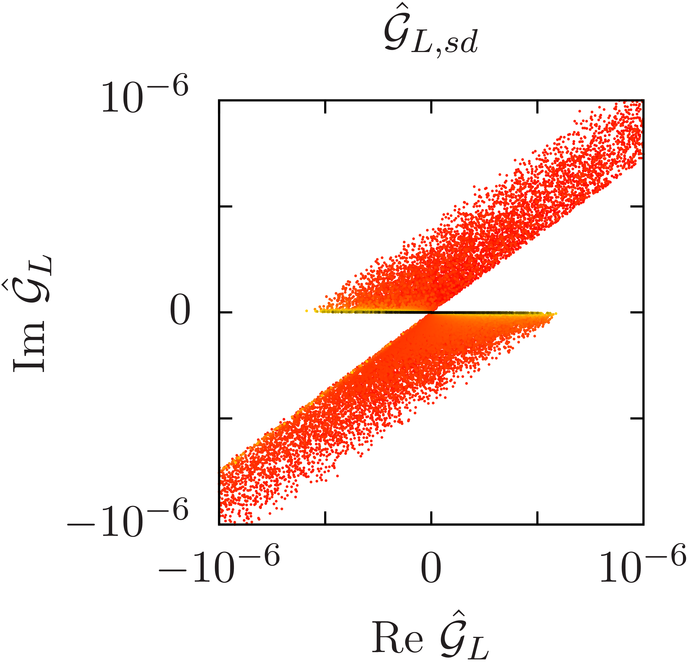}\qquad\includegraphics[width=7cm]{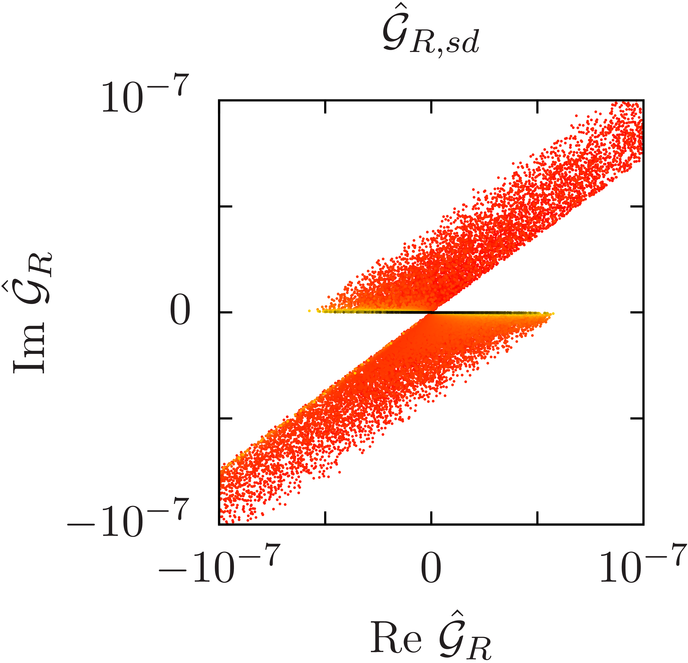}\\[1em]
    \includegraphics[width=7cm]{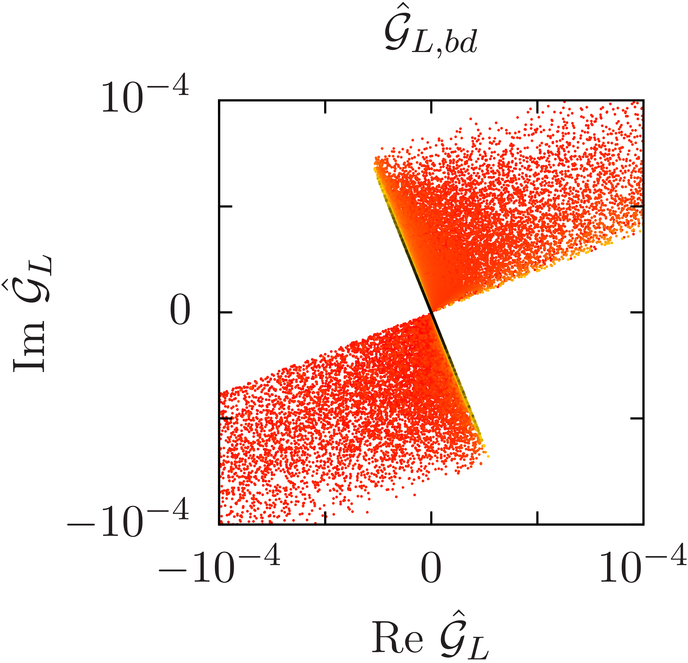}\qquad\includegraphics[width=7cm]{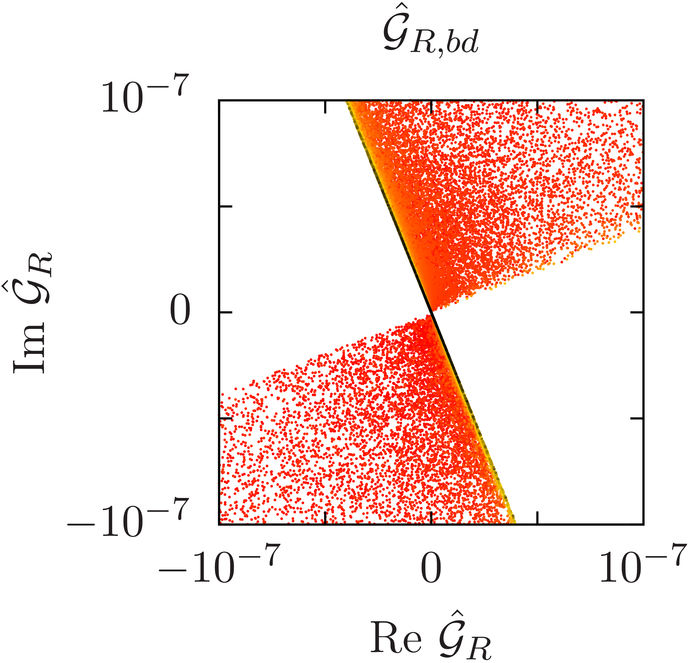}\\[1em]
    \includegraphics[width=7cm]{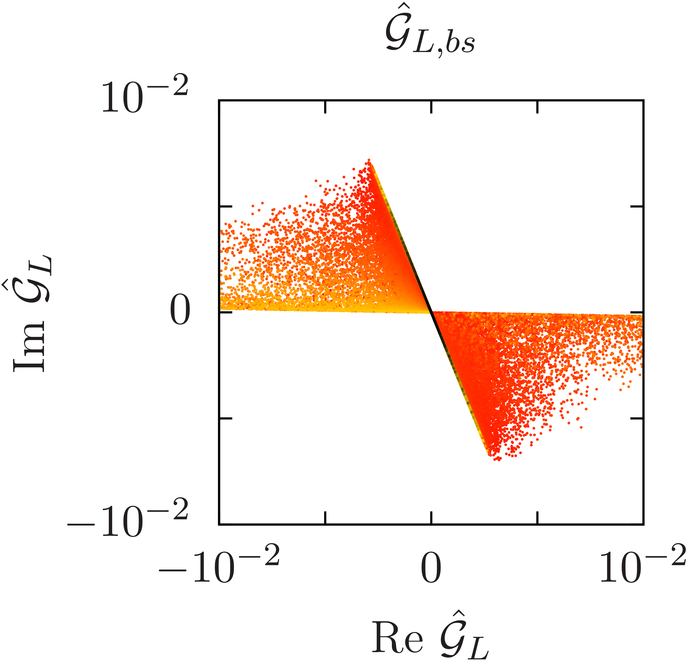}\qquad\includegraphics[width=7cm]{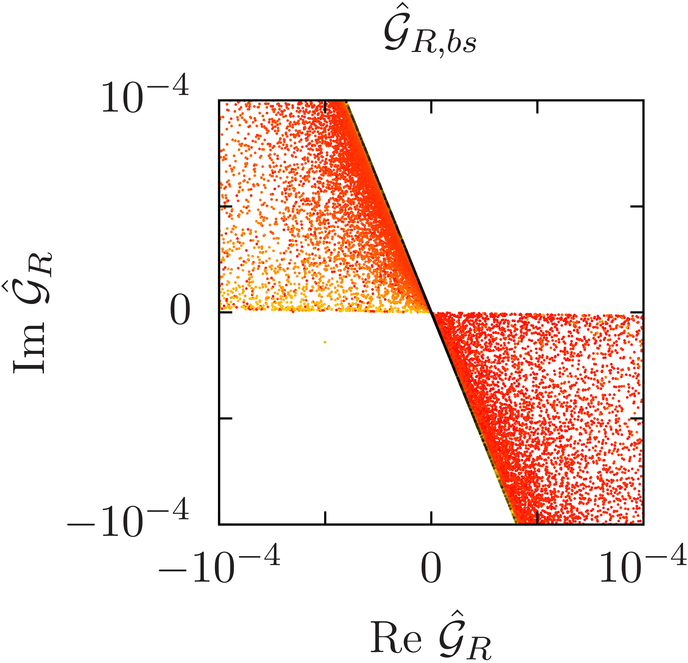}
  \end{center}
  \caption{\it The couplings of the lightest flavour gauge boson to the down-type SM quarks.
  \label{fig:couplings24_down}}
\end{figure}

\begin{figure}
  \begin{center}
    \includegraphics[width=7cm]{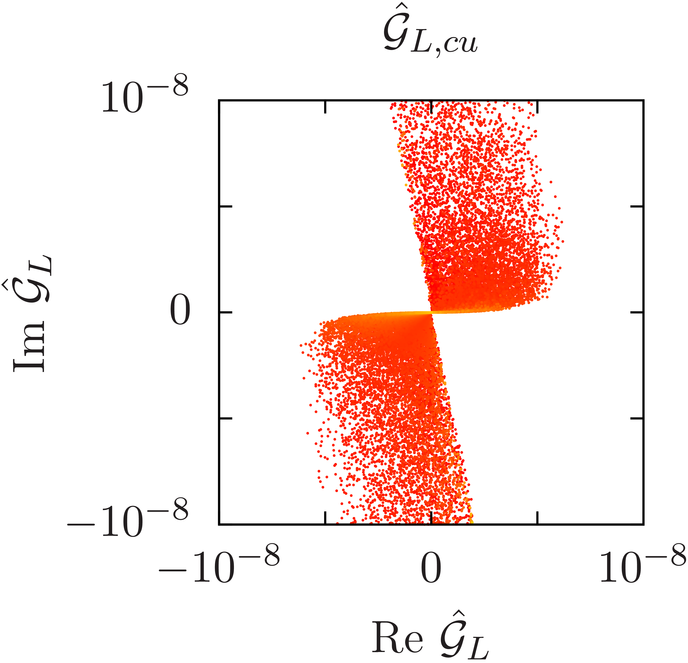}\qquad\quad\includegraphics[width=7cm]{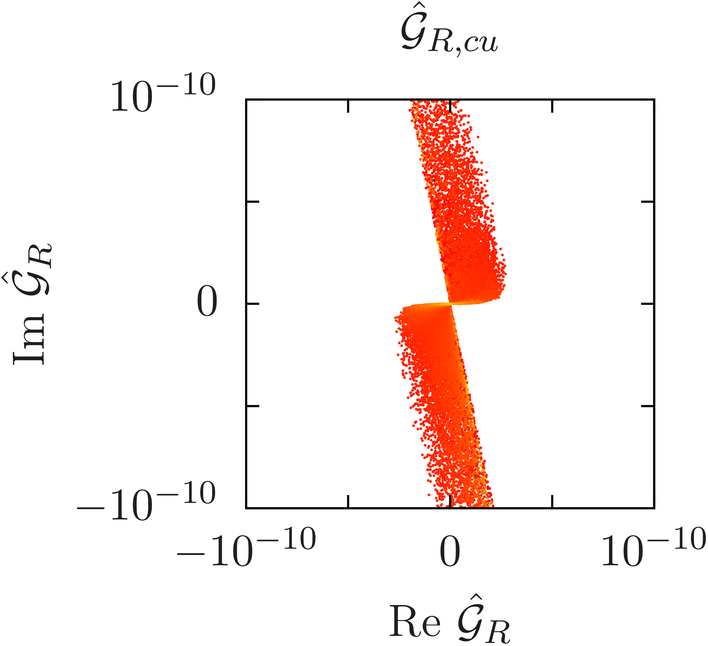}\\[1em]
    \includegraphics[width=7cm]{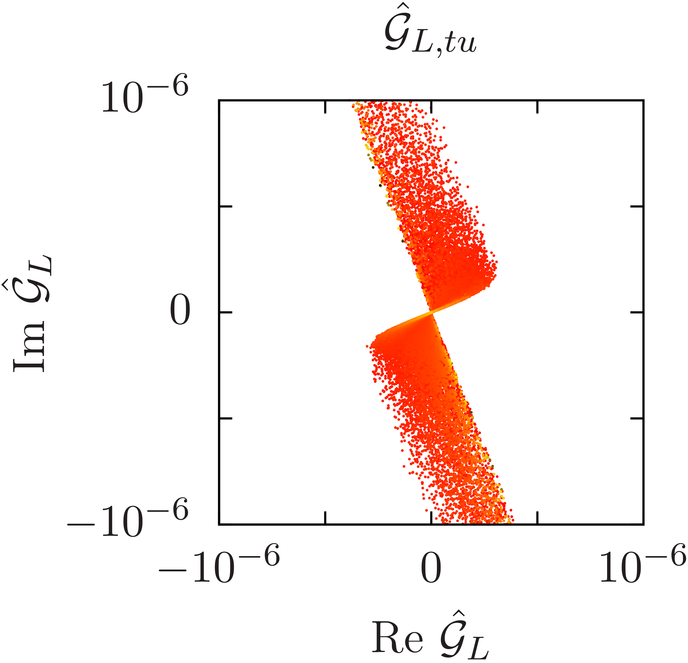}\qquad\quad\includegraphics[width=7cm]{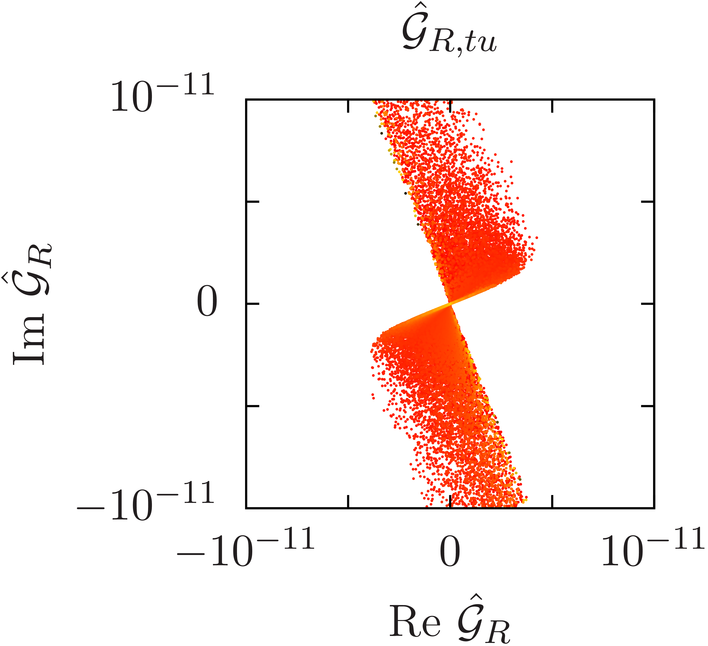}\\[1em]
    \includegraphics[width=7cm]{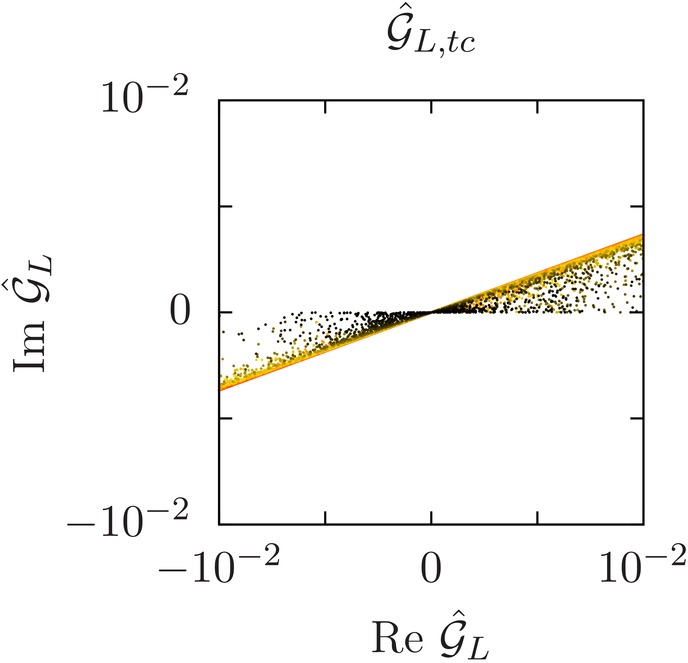}\qquad\quad\includegraphics[width=7cm]{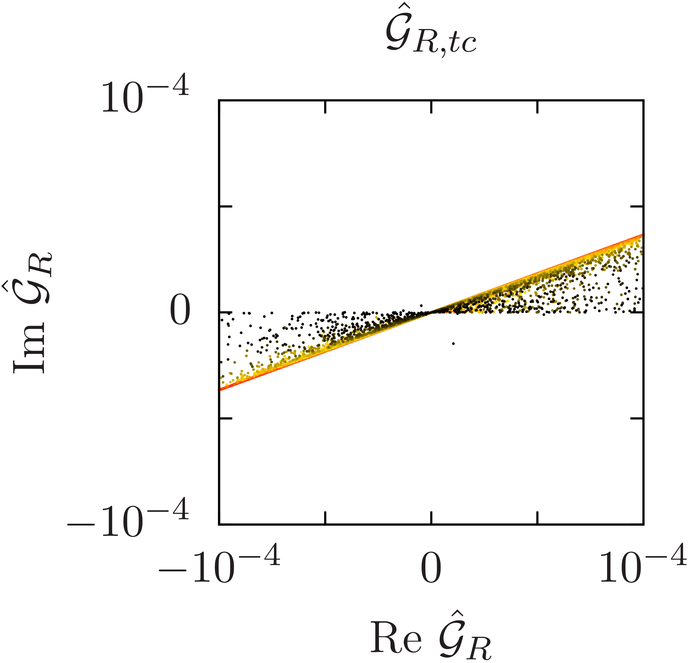}
  \end{center}
  \caption{\it The couplings of the lightest flavour gauge boson to the up-type SM quarks.
  \label{fig:couplings24_up}}
\end{figure}

As an example, we report in fig.~\ref{fig:couplings24_down} and \ref{fig:couplings24_up}
the couplings of the lightest gauge boson to the light down- and up-type quarks for the exclusive
$V_{ub}$ case and scattering on all the parameters of the model. The colour-coding
corresponds to the mass splitting of the lightest to the next-to-lightest flavour
gauge boson mass. Namely, red points correspond to large splitting while black points to
degeneracy, yellow to intermediate splitting.
We observe that the flavour-violating couplings with the last (first) two generations are
in general the largest (smallest) ones. This is related to the sequential breaking of
the flavour symmetry encoded into hierarchical structure of the flavon VEVs.

\clearpage{\cleardoublepage}
%
%

\phantomsection
\addcontentsline{toc}{section}{References}
\providecommand{\href}[2]{#2}\begingroup\raggedright\endgroup

\end{document}